\pdfoutput=1
\documentclass[usenatbib,fleqn]{mnras}
\usepackage{ifthen} \newboolean{pubformat} \setboolean{pubformat}{true}
\usepackage{natbib}		
\usepackage[draft]{pdfpages}
\usepackage{amsmath}
\usepackage{hyperref}  

\usepackage{subfig}
\usepackage{graphicx}
\usepackage[countmax]{subfloat}  

\newcommand{\continuefloat}
{\ifthenelse{\boolean{pubformat}}{\contcaption{}}{\ContinuedFloat}}
\ifthenelse{\boolean{pubformat}}
{
\usepackage{fixltx2e}     
}{
\usepackage{times}
}
\newcommand{\onecol}[1]{\includegraphics[width=88mm]{#1}}
\newcommand{\twocol}[1]{\includegraphics[width=176mm]{#1}}
\usepackage{graphicx}
\hypersetup{
    colorlinks=true,
    linkcolor=blue,
    filecolor=magenta,      
    urlcolor=cyan,
}
\title[The Orphan Stream in \Gaia\ DR2]
{The course of the Orphan Stream in the Northern Galactic Hemisphere traced with \Gaia\ DR2}
\author[M. A.\ Fardal et al.]   
{Mark A.\ Fardal$^{1,2}$\thanks{E-mail: fardal@stsci.edu}, 
Roeland P. van der Marel$^{1,3}$,
Sangmo Tony Sohn$^1$,\newauthor
Andres del Pino Molina$^1$\\
$^1$Space Telescope Science Institute, 3700 San Martin Drive, Baltimore, MD 21218, USA\\
$^2$Department of Astronomy, University of Massachusetts, Amherst, MA 01003-9305, USA\\
$^3$Center for Astrophysical Sciences, Department of Physics \& Astronomy,
Johns Hopkins University, Baltimore, MD 21218, USA
}  
\date{Submitted to MNRAS \today}
\pagerange{\pageref{firstpage}--\pageref{lastpage}} \pubyear{2018} 
\ifthenelse{\boolean{pubformat}}
{
\voffset -1.4cm    
\setlength{\textheight}{241mm}  
}{
\setlength{\textwidth}{6.5in}
\setlength{\oddsidemargin}{0in}
\setlength{\evensidemargin}{0in}
\setlength{\textheight}{227.6mm}  
\topmargin=0mm
\headheight=0mm
\headsep=0mm
\footskip=13mm
}
\renewcommand{\sun}{\odot}  

\newcommand{\Gyr}{\,\mbox{Gyr}}

\newcommand{\kpc}{\,\mbox{kpc}}

\newcommand{\kms}{\,\mbox{km}\,\mbox{s}^{-1}}
\newcommand{\masyr}{\,\mbox{mas}\,\mbox{yr}^{-1}}

\newcommand{\msun}{\,M_{\sun}}
\newcommand{\degree}{\degr}
\newcommand{\pmlon}{\mu_\Lambda}
\newcommand{\pmlat}{\mu_B}
\newcommand{\nbody}{$N$-body}
\newcommand{\tsim}{\sim\!}
\newcommand{\feh}{[\mathrm{Fe}/\mathrm{H}]}
\newcommand{\afe}{[\alpha/\mbox{Fe}]}

\newcommand{\lnorm}{\Lambda_{100}}
\newcommand{\gtrsim}{\ga}
\newcommand{\ltrsim}{\la}
\newcommand{\Gaia}{{\it Gaia}}
\newcommand{\HST}{{\it HST}}
\defcitealias{sesar17}{S17}
\defcitealias{law10}{LM10}
\defcitealias{dierickx17}{D17}
\hypersetup{draft}   
\begin{document}
\maketitle    
\label{firstpage}
\begin{abstract}
The Orphan Stream is one of the most prominent tidal streams in the Galactic halo.
Using data on red giants, RR Lyrae, and horizontal branch stars
from \Gaia\ and other surveys, we determine the proper motion of the
Orphan Stream over a path of more than $90 \degree$ on the sky.  We also
provide updated tracks for the sky position, distance, and
radial velocity of the stream.  Our tracks in these latter dimensions
mostly agree with previous results.  However, there are significant
corrections to the earlier distance and latitude tracks as the stream
approaches the Galactic disk.
Stream stars selected with three-dimensional kinematics
display a very tight red giant sequence.
Concordantly, we find that applying a proper motion cut removes the most
metal-rich stars from earlier spectroscopic samples of stream stars,
though a significant dispersion remains indicating a dwarf galaxy origin.
The deceleration of the stream towards its leading end suggests a circular
velocity of $\tsim 200 \kms$
at a galactocentric radius $\tsim 30 \kpc$, consistent with other
independent evidence.
However, the track of the stream departs significantly from an orbit;
the spatial track does not point along the same direction
as the velocity vector, and it exhibits a lateral wiggle that is unlikely
to match any reasonable orbit.
The low metallicity and small dispersion of the stream in the various coordinates
point to a progenitor with a relatively low dynamical mass $\tsim 10^8 \msun$.  
\end{abstract}
\section{INTRODUCTION}
The Galactic halo is populated by many stellar inhomogeneities,
including galaxies, clusters, streams, shells, and clouds.
These serve as a record of the Galaxy's accretion history.
For many years astronomers have also hoped that the better-defined tidal
streams will also serve as a probe of the Galaxy's gravitational potential.
The translation from observed stream properties to gravitational
potential is not trivial,
but some attempts in this direction have been made with the 
Sagittarius, GD-1, Pal 5, and Orphan streams.
Even in the most prominent streams it is often difficult to disentangle
stream and unrelated stars,
which complicates measurement of the bulk properties of the stream.
Furthermore, the proper motion of halo stars has been difficult to measure,
so in most cases two of the six dimensions of phase space are missing.
The second release of data (DR2) from the \Gaia\ mission \citep{gaia16,gaia18}
promises to help immensely with these difficulties, since it provides
data with accurate astrometric parameters and photometry for about two billion stars
spread over the entire sky.

The Orphan Stream is one of the most prominent features in the Galactic halo,
extending over $90 \degree$ in length with a width of only 1--$2 \degree$.
The stream was initially traced in turnoff stars in Sloan Digital Sky Survey (SDSS) data
\citep{belokurov06,grillmair06,belokurov07},
and named for its lack of an obvious progenitor.
In an impressive piece of detective work, \citep{newberg10} traced its path in 
sky position, distance, and velocity coordinates
using blue horizontal branch (BHB) stars from SDSS,
and obtained the first reasonably accurate orbit and \nbody\ models of the stream.  
Knowledge of the stream's stellar content was increased using RR Lyrae (RRL) stars
\citep{sesar13,hendel18} and red giants \citep{casey13}.
\citet{sohn16} measured its proper motion in two
{\it Hubble Space Telescope} (\HST) fields.  
\citet{grillmair15} used Dark Energy Camera observations
to trace the stream further south as it nears the Galactic disk.
However, the origin and total extent of the stream are still uncertain,
and its implications for the Galactic potential are still largely unclear.

In Section~\ref{sec.analysis} of this paper, we use data from \Gaia\ DR2 to measure the proper motion
along more than $90 \degree$ of the Orphan Stream.
We also revise previous estimates for the sky path, distance, radial
velocity, and stellar population of the stream.
In Section~\ref{sec.interpretation}, we briefly examine the derived stream track and compare it to
simple orbital models for the stream.
Section~\ref{sec.conclusions} presents our conclusions.
\section{DATA ANALYSIS}
\label{sec.analysis}
\subsection{Overview of method and previous results}
\label{sec.method}
The extent of the Orphan Stream on the sky is
best delineated at present in maps made
from the abundant stars on the upper main sequence and turnoff
\citep[e.g.,][]{belokurov06,grillmair06,belokurov07,newberg10,grillmair15}.
The stream runs roughly north-south over a length $\gtrsim 90 \degree$, 
while it is only $\tsim 1\degree$--$2\degree$ wide.
We adopt the stream coordinate system with longitude and latitude $\Lambda$, $B$
introduced by \citet{newberg10}.
In this system, the stream remains within a few degrees of
the equator line $B = 0$ over the longitude
range $-35 \degree \ltrsim \Lambda \ltrsim 60 \degree$.
The exact path of the stream is uncertain at both ends of the observed range.
In the north this is due both
to a real decrease in the stellar density of the stream,
and to the increasing distance which makes detection more difficult \citep{newberg10}.
In the southern region, this is mainly due to increasing contamination from the Milky Way disk
as well as uncertainty in the extinction correction \citep{grillmair15}.
From its geometry and radial velocity, the stream is known to
flow northwards \citep{newberg10}, in the direction of decreasing $\Lambda$.  

Rather than recalculating these spatial maps, we will use summary
information about them such as the central track and stream width.
\Gaia\ contributes little useful information at magnitudes corresponding
to the stream's main-sequence turnoff.  Instead, we will use \Gaia\ to
gain new information on the evolved stars in the stream.
We do not search for a continuation of the stream beyond the region
where it is currently known to exist, leaving that for future work.

We study the stream
in the sky region $-35 \degree < \Lambda < 75 \degree$,
$-6 \degree < B < 6 \degree$.
Most of the previous knowledge of the Orphan Stream was obtained with
photometry and spectroscopy from SDSS.
We instead use the Pan-Starrs1 (PS1) survey \citep{chambers16} as our preferred source of photometry,
because its coverage extends further south 
to a limit of $\delta > -30 \degree$,
or $\Lambda \approx 51 \degree$ near the stream track.
We split the region under consideration into a
northern region $\Lambda < 50\degree$ 
where we use PS1 data, and a southern one $\Lambda > 50\degree$
where only \Gaia\ photometry is publicly available.  

We obtain \Gaia\ DR2 data from the ESAC archive, using the PS1 cross-matched table for
the northern region and the standard \Gaia-only source table for the southern one.
We correct for extinction with the dust maps of \citet{schlegel98}
accessed with the python code {\tt sfdmap}.
We use the rescaled dust extinction coefficients from \citet{schlafly11}
for PS1 photometry,
and those from \citet{malhan18} for \Gaia\ photometry.

We use sharp selection cuts rather than probabilistic methods
to define the stream sample.
We begin by selecting all stars with valid parallax and proper motion values.
To remove foreground stars, we use only stars with
parallax $\varpi$ within $2 \sigma_\varpi$ of the value
expected for stream stars.  
The expected $\varpi$ combines the true parallax,
as computed from the distance track of \citet{newberg10}
with the small average DR2 parallax bias 
$\varpi_0 = -0.03 \masyr$ found by \citealp{lindegren18},
though these terms are both small corrections compared to $\sigma_\varpi$.
We transform the proper motion values from the equatorial $(\alpha, \delta)$
coordinates used by \Gaia\
to stream-aligned coordinates
$\pmlon$ and $\pmlat$, where $\pmlon \equiv \cos B \, d\Lambda/dt$
(i.e., this is a physical and not a coordinate angular speed).  

We also apply some quality cuts to the \Gaia\ sources.
One cut is based on the flux excess factor $E \equiv \mbox{\tt
phot\_bp\_rp\_excess\_factor}$, which compares the $G$ magnitude to
the value expected from the $G_{BP}$ and $G_{RP}$ magnitudes.
We used a slightly relaxed version of the cut in equation C.2 of
\citet{lindegren18}, namely
$1 + 0.015 (G_{BP} - G_{RP})^2 < E < 1.5[1.3 + 0.06 (G_{BP} - G_{RP})^2]$.
Another cut removes sources with bad astrometric fits
following equation~C.1 in \citet{lindegren18}:
defining $u \equiv (\mbox{\tt astrometric\_chi2\_al} /$ $\mbox{\tt astrometric\_n\_good\_obs\_al} - 5)^{1/2}$,
we require $u < 1.2 \times \max(1, \exp(-0.2(G-19.5)))$.
Another simply excludes stars with proper motion errors in RA or declination
of $> 1 \masyr$, as these are large enough to make the proper motion selection
cuts we use unreliable.

Our method for detecting and describing the stream is iterative.
We start from a good guess for the location of the stream in the various
observables, search for the presence of a clump of stars,
and refine our initial guess.
We define an off-stream sample made up of stars outside our latitude cut,
and verify that the on-stream clump is absent from the off-stream sample
to make sure it represents the stream and not some other substructure.
We then cycle through the various observables and
datasets, gradually extending and refining our best-fit stream track.
Implicit in our method is the assumption that the stream can be modeled
by a single track, with a dispersion about the track and a stellar population
that does not change too radically with position.
The text will describe how we establish tracks for each observed dimension
starting from previously published estimates of the spatial, distance, and
velocity tracks.
Rather than giving results from each stage of the iteration, however,
the plots, statistics, and formulae we present are all based on the final 
estimate of the stream track in each dimension.

In the northern region we will also
constrain the stream with spectroscopy from SDSS,
and the catalog of RRL in PS1 from \citet{sesar17}.
As initial guesses for the stream location, we will use the 
smooth empirical tracks (rather than orbits)
from \citet{newberg10} and \citet{grillmair15}.
We will also compare our results to datasets including
the survey of Orphan Stream RRL stars with radial velocity by \citet{sesar13},
the survey of stream RGB candidate members by \citet{casey13},
and the proper motion measurements using \HST\ by \citet{sohn16}.
\subsection{Stream distance from Blue Horizontal Branch and RR Lyrae stars}
\label{sec.distance}
\begin{figure*}
\subfloat[]{\includegraphics[width=60mm]{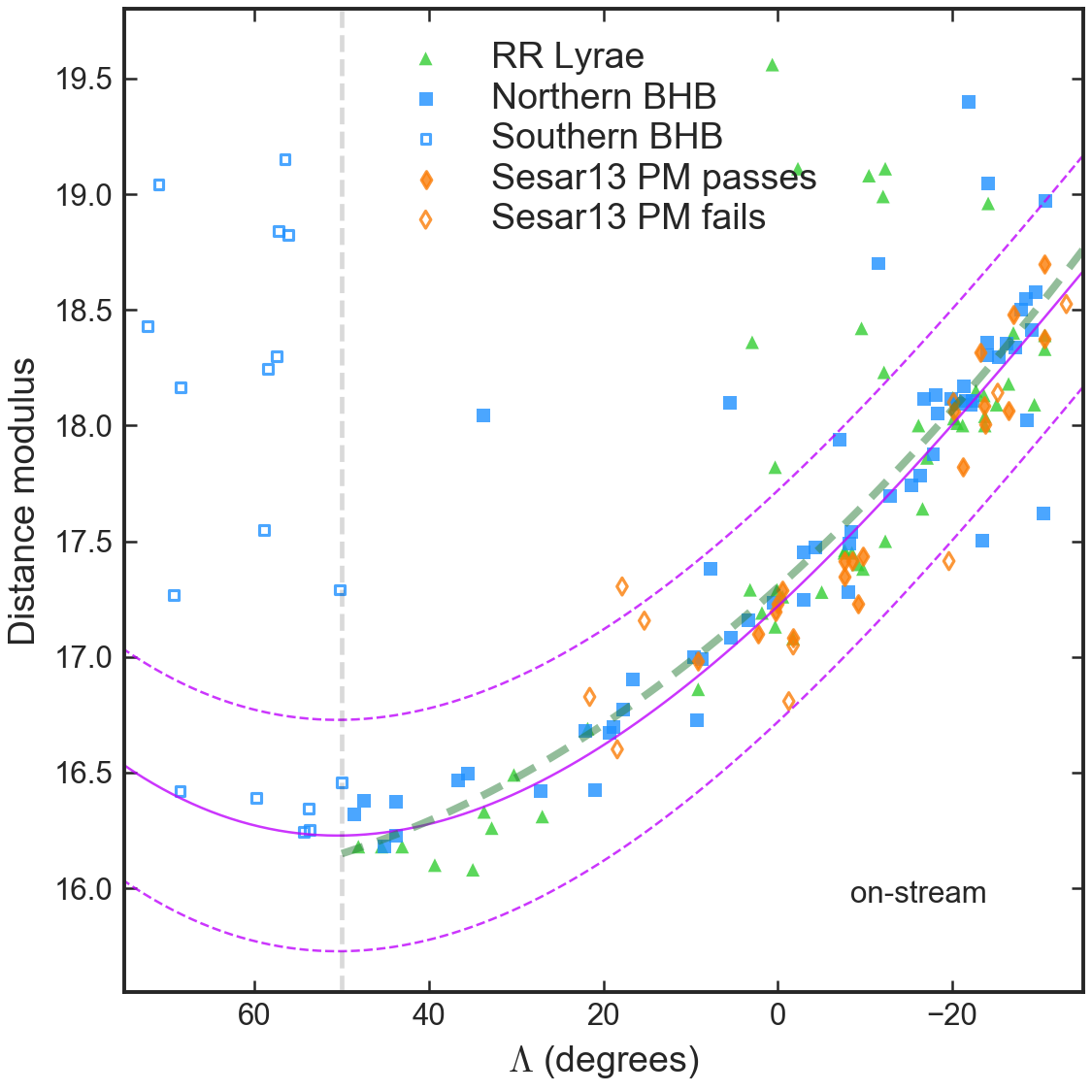}}
\subfloat[]{\includegraphics[width=60mm]{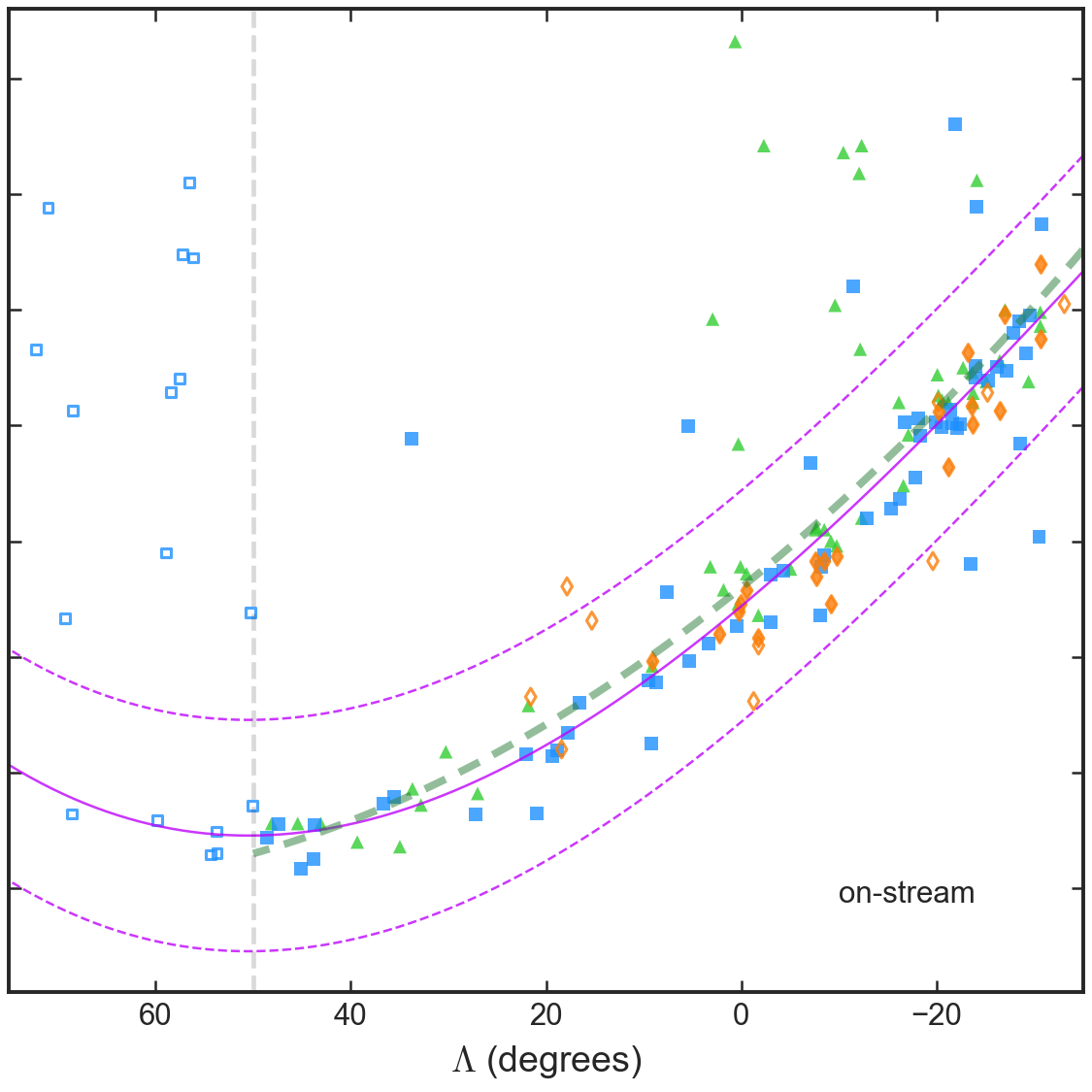}}
\subfloat[]{\includegraphics[width=60mm]{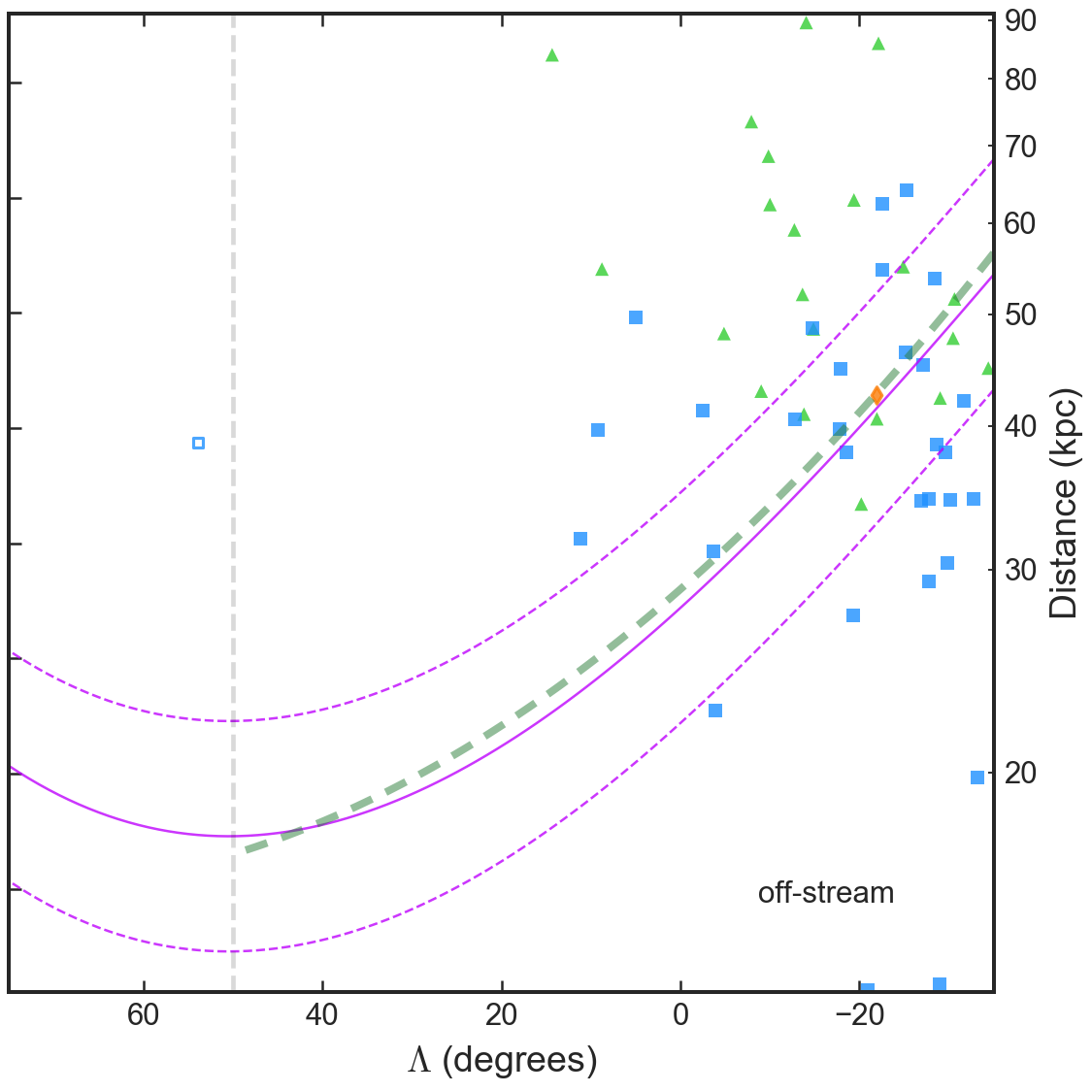}}
\caption{
\label{fig.distance}
Panel (a): distance modulus to BHB and RRL stars within the on-stream sky region,
defined as a latitude interval $\pm 2.25 \degree$ around our adopted spatial track.
This panel uses our original distance calibrations, for which RRL stars
have systematically lower distances than the BHBs.
Proper motion and photometric cuts are applied as discussed in the text.
Parallax and data quality cuts are also applied throughout the paper.
The boundary of the PS1 survey (vertical line) lies at $\Lambda = 50\degree$.
RRL stars are taken from the PS1 sample \citep{sesar17} and so are present
only in the northern region ($\Lambda < 50\degree$).
In the southern region ($\Lambda > 50\degree$), the BHB selection method 
uses only \Gaia\ photometry which is less effective at screening contaminants.
High-probability stream members from the separate RRL survey of \citet{sesar13} are also shown,
with point style indicating whether they
pass or fail our proper motion cut as indicated by the legend.
Stars concentrate around the distance track
of Equation~\ref{eqn.distancetrack}, shown by the solid line.
Short-dashed lines show the limits we use for distance modulus selection.
The long-dashed line shows the empirical distance track of \citet{newberg10}. 
Panel (b): Same, but using our revised BHB and RRL distance calibrations.
The BHB and RRL stars along the stream now agree better on average.
Panel (c): Same as (b), but for the off-stream sky region.
Here the concentration along the stream track is absent.
}
\end{figure*}
BHB stars are excellent distance tracers for metal-poor objects such as
the Orphan Stream.
While BHB stars are bluer than the main-sequence turnoff and thus less
contaminated than redder stars, 
selection via a single color still mixes in unrelated objects such as
blue stragglers, white dwarfs, and quasars.
This contamination can be
reduced using multiple colors, either in the mid-infrared
or using surface-gravity-sensitive colors such as the SDSS $u$-band or
the PS1 $z$-band.
\citet{vickers12} proposed a BHB selection method making use of
the $g$, $r$, $i$, and $z$ bands
to screen out these contaminants.
However, we found these relatively stringent cuts were not well suited
to the increased photometric errors as the stream approaches distances of $\tsim 50 \kpc$.
To some extent, we can rely on the proper motion to screen contaminants
from our stream sample.  
We will therefore use a more relaxed selection boundary than
\citet{vickers12}, defined as follows:
\begin{gather}
-0.3 < g-r < 0 \notag \\
-0.062 + 0.48 \, (g-r) < i-z < -0.02 \notag \\
-0.30 + 1.96 \, (g-r) < g-z < 0 + 1.96 \, (g-r) \; .
\end{gather}
In the southern region we have only \Gaia\ photometry which essentially provides
a single color.
Using old, metal-poor horizontal-branch stars in the MIST isochrone set \citep{dotter16,choi16} as a guide,
we approximate the relationship between $G_{BP}-G_{RP}$ and $g-r$ for these stars as
\begin{equation}
g-r \approx -0.26 + 0.72 \, (G_{BP}-G_{RP}) \; ,
\end{equation}
and then use the $g-r$ cut just stated to obtain
a \Gaia\ color cut.  
Compared to the more elaborate PS1 selection this allows more contaminants,
particularly blue stragglers, but we show below
that we still can detect the stream in the region of interest.

RR Lyrae stars are also excellent distance tracers, which are in a similar evolutionary
stage and have similar luminosities to BHB stars.
With a sufficient number of observations, RRL samples
can have extremely high purity due to their distinctive light curves.
The PS1 RRL sample of \citet{sesar17} is a large-scale homogeneous sample where
distances are derived from the brightness and time variation in multiple bands.
We cross-match this sample to \Gaia\ DR2.
We restrict the sample to stars with tabulated RRab score $> 0.8$, to avoid mixing
in RRc stars and other contaminants.

To guide our analysis, we also use the RR Lyrae radial velocity survey of \citet{sesar13}
that specifically targeted the Orphan Stream.  Candidate RR Lyrae stars in
the vicinity of the stream were obtained
from several time-series surveys and observed spectroscopically,
taking care to correct effects of pulsation phase on the velocity.
The sample cross-matched to \Gaia\ DR2 contains 50 total stars
spanning the range $-48 \degree < \Lambda < 22 \degree$.
\citet{sesar13} mark 31 of these as likely stream members based on
comparison with distance and velocity tracks from \citet{newberg10}.
The method and data used to compute distance moduli
differs between \citet{sesar13} and \citet{sesar17},
but the rms difference is only $0.12$ mag.  This level of disagreement
is consistent with both samples having independent random distance errors
of a mere $\tsim 4\%$.

\begin{figure*}
\includegraphics[width=127mm]{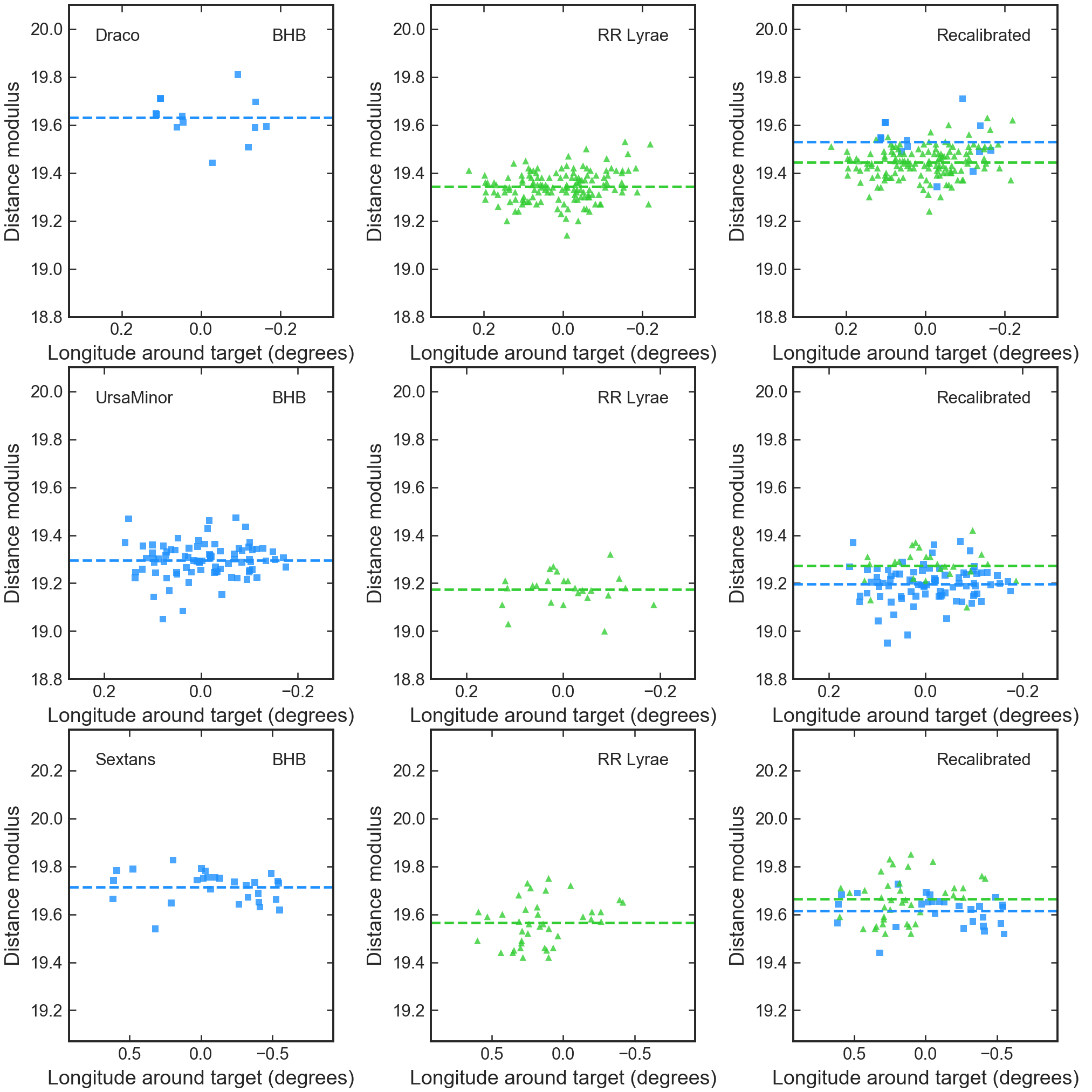}
\caption{
\label{fig.dsph}
Distance modulus to BHB and RRL stars in dSph galaxies.
The rows show results for Draco, Ursa Minor, and Sextans respectively.
The horizontal axis shows one sky coordinate
and is used merely to separate the points.
Each panel uses the same vertical scale.
The dashed lines show the median distance modulus for each type.
Left column: distance modulus for BHB stars
is computed from PS1 photometry with the \citet{deason11b} calibration.
Center column: distance modulus for RRL stars
is taken from the table of \citet{sesar17}.
Right column: the same results for both types of star, but
using our revised calibration (see Section~\ref{sec.distance}).
}
\end{figure*}

We initially estimate absolute magnitudes and distance moduli
for the BHB stars using the color-based formula of \citet{deason11b}.
We make a minor correction from SDSS bands ($g_S, r_S$)
to PS1 bands ($g, r$), again using linear relations that approximate colors of 
horizontal branch stars in the MIST isochrone set: 
\begin{gather}
g_S - r_S = 0.014 + 1.114 (g-r) \notag \\
M'_\mathit{g_S,BHB}(g_S-r_S) = 0.434 - 0.169 (g_S - r_S)  + 2.319 (g_S - r_S)^2 \notag \\
  \hspace{4em} + 20.449 (g_S - r_S)^3 + 94.517 (g_S - r_S)^4 \notag \\
M'_\mathit{g,BHB} = M'_\mathit{g_S,BHB}(g_S-r_S) - 0.014 - 0.120 (g - r) 
\end{gather}
Here the prime on the mean BHB absolute magnitude $M'_\mathit{g,BHB}$ 
indicates it is an initial calibration that we will adjust later.
For the southern region where only \Gaia\ photometry is available,
we use approximate transformations between \Gaia\ and SDSS bands
(again based on MIST isochrones) to apply the relation above:
\begin{gather}
g_S - r_S = -0.27 + 0.80 (G_{BP}-G_{RP}) \notag \\
g_S - G = - 0.15 + 0.60 (G_{BP}-G_{RP}) \notag \\
M'_\mathit{G,BHB} = M'_\mathit{g_S,BHB}(g_S-r_S) - (g_S-G) 
\end{gather}

We use an initial guess for the stream latitude that closely follows
the results of \citet{newberg10} and \citet{grillmair15}.
Our spatial selection uses stars within $2.25 \degree$ of this track.
Assuming the $0.9 \degree$ latitude dispersion estimated by \citet{belokurov06},
this is approximately $2.5 \sigma$.
We also initially require the estimated distance modulus to lie within
0.5 mag of the track of \citet{newberg10}.
Binning the stars in $10 \degree$ intervals of $\Lambda$, we find
a clump of stars lying at high $\pmlon$ for $\Lambda \sim 50\degree$.
This clump can be followed over the entire
range of $\Lambda$ in the northern sample, though it is less distinct
from the background population at low $\Lambda$.
Fitting the values found from the bins,
we obtain an initial smooth track for the stream's proper motion.
When the distance moduli of 
stars near this proper motion track are plotted versus $\Lambda$,
a distinct stream is apparent.
Figure~\ref{fig.distance} shows the BHB stars as cyan points in the left panel,
with solid symbols denoting the northern sample using PS1
photometry and open symbols the southern sample.
In the figure we use our final stream proper motion and position tracks to define
the sample, in order to present only our final converged results.
A similar concentration is not seen in the sample of stars outside our latitude cut.

We then apply the same position and proper motion cuts to the RR Lyrae sample,
in which a clump can be found in proper motion space following the same trend
as the BHB stars.
Using the default distance modulus values from \citet{sesar17},
this yields a similar concentration along the stream distance track
(green points in left panel of Figure~\ref{fig.distance}).
Thus we have a clear detection of the Orphan Stream in the \Gaia\ data.

In panel (a) of Figure~\ref{fig.distance} there appears to be an offset between the
BHB and RR Lyrae stars, with the latter apparently nearer on average.
(A similar offset is suggested by figure 1 of \citealp{sesar13}, with the 
majority of the stream RR Lyrae on the near side
of the orbital tracks from \citealp{newberg10}.)
It seems unlikely that these two types of stars could consistently lie at
different distances over such a long stretch of the stream,
whereas a mismatch of our distance scales would not be particularly surprising.
There are several places where systematic error could arise,
including our translation from the SDSS-based distance scale
to PS1 and \Gaia\ for the BHB stars.
Both the \citet{deason11b} BHB and \citet{sesar17}
RR Lyrae magnitudes are calibrated against globular clusters.
It is likely that both calibrations should be affected by metallicity,
but neither sample takes this into account explicitly.
Furthermore, the relationship between BHB and RRL stars might be different
in globular clusters vs dwarf galaxies, and Orphan appears to be a remnant
of the latter class.
In dwarf galaxies one might expect that
whether a star winds up as a BHB or an RRL is primarily due to
metallicity. A globular, in contrast, typically has a very small metallicity
($\feh$) range, but also has sub-populations with inferred differences in helium
abundances and relative metal abundance patterns.
These differences, rather than just the value of $\feh$,
are likely to determine whether a star winds up as a BHB or an RRL.
It thus seems plausible that the BHB and RRL distance scales
are not completely consistent for the stream sample here.

We therefore recalibrate the BHB and RRL distances
using three dSph galaxies with metallicities similar to that
of the Orphan Stream: Draco, Ursa Minor, and Sextans.
With our original calibrations, we find in each case
an offset between BHB and RRL stars in the same direction as for the Orphan Stream
(Figure~\ref{fig.dsph}), although its size seems to vary a bit.
The mean offset in these three galaxies is 0.19~mag.  To reconcile the
distance scales, we simply split the difference.
We add 0.1~mag to the absolute magnitude
formula of \citet{deason11b}:
implying a distance modulus 0.1 mag smaller.
\begin{gather}
M_\mathit{g,BHB} = M'_\mathit{g,BHB} + 0.10  \notag \\
M_\mathit{G,BHB} = M'_\mathit{G,BHB} + 0.10 
\end{gather}
We also subtract 0.1 mag from the PS1 RRL absolute magnitudes, 
or equivalently assume a distance modulus 0.1 mag larger than given in the original table.
The recalibrated distance moduli for the Orphan sample
are shown in the panel (b) of Figure~\ref{fig.distance}.
Panel (c) of Figure~\ref{fig.distance} shows the BHB and RRL in the off-stream region.
The strong concentration along the stream distance track is absent in this plot,
confirming that it originates from the Orphan Stream.

Of course, given that the distance scales for both types of object
are uncertain, our newly reconciled distance scale is also uncertain.
For an additional test we matched the RRL passing all of our sample cuts to the sample of
\citet{hendel18}.  This sample uses near-infrared photometry of the
stars in \citet{sesar13} and thus arguably
has better distance estimates than \citet{sesar17}, 
though sources of systematic error remain.
We find the median difference in distance modulus between our
recalibrated values and those of \citet{hendel18} is $0.04$ mag, or a distance
offset of only 2\%.  This is probably within the level of systematic error
for the stellar tracers used here, and 
we regard it as acceptable agreement.

The distance modulus of the stream appears well described by the curve
\begin{gather}
\mathit{DM}(\Lambda) = 17.22 - 3.52 \, \lnorm + 2.29 \, \lnorm^2 + 1.58 \, \lnorm^3 \notag \\
\lnorm \equiv \Lambda / 100\degree
\label{eqn.distancetrack}
\end{gather}
The track is in reasonable agreement with the empirical track of \citet{newberg10}
shown with the long-dashed line, but puts the stream slightly closer for much of the
observed range.  Furthermore, our track flattens at $\Lambda=50\degree$
above which the distance rises again, though the exact form of this
turn-up is uncertain due to the small number of stars constraining it.

The dispersion in magnitude around this track appears to be
roughly 0.2 mag for RRL stars, and 0.3 mag for BHB stars.
This implies an upper limit on the overall distance dispersion of $\approx 10$--15\%.
These dispersions most likely stem from the combined effects of 
sample contamination and intrinsic dispersion in the source brightnesses,
rather than the intrinsic thickness of the stream.
Using near-infrared photometry,
\citet{hendel18} formally estimated a distance dispersion of 0.22 mag
around their orbital track.
We note that in that paper as well as this one,
much of the overall dispersion appears to originate
at the northern end of the stream, whereas the portion in
$0\degree < \Lambda < 50\degree$ is much better collimated.  

Figure~\ref{fig.distance} also shows for comparison the
28 high-confidence stream RRL from \citet{sesar13}
that fall within our latitude cut, with closed (open) symbols
denoting those that pass (fail) our proper motion cut.
Although most of these 28 stars pass our cut, 10 do not.
The 7 stars labeled RR19, RR30, RR31, RR43, RR46, RR47, RR49
are offset from our trend by more than $1 \masyr$ in either
longitude or latitude directions, compared to typical errors
of $0.3 \masyr$.  
\citet{hendel18} previously noted the tension between
\Gaia\ measurements and expectations for five of these stars,
and furthermore found that another (RR19)
has a light curve inconsistent with a genuine RRL star.
The stars failing the proper motion cut are outliers in several
other ways.  
They include the four at highest $\Lambda$, in the
Galactic longitude range $245\degree < l < 258\degree$
in figure 2 in \citet{sesar13}.
They also include the four ``stream'' stars most discrepant from our distance tracks.
Finally, they also include the two stars with the highest metallicity in the
``stream'' sample (RR46 and RR47).
\citet{sesar13} noted both a significant
metallicity dispersion and significant gradient of spectroscopic metallicity
along the stream.
Removal of these stars would significantly suppress both of these properties. 
It would however not eliminate either entirely, which is
significant insofar as it relates to the nature of the stream's progenitor.
\subsection{Velocity and stellar population from spectroscopic sample}
\label{sec.cmd}
\begin{figure*}
\includegraphics[width=72mm]{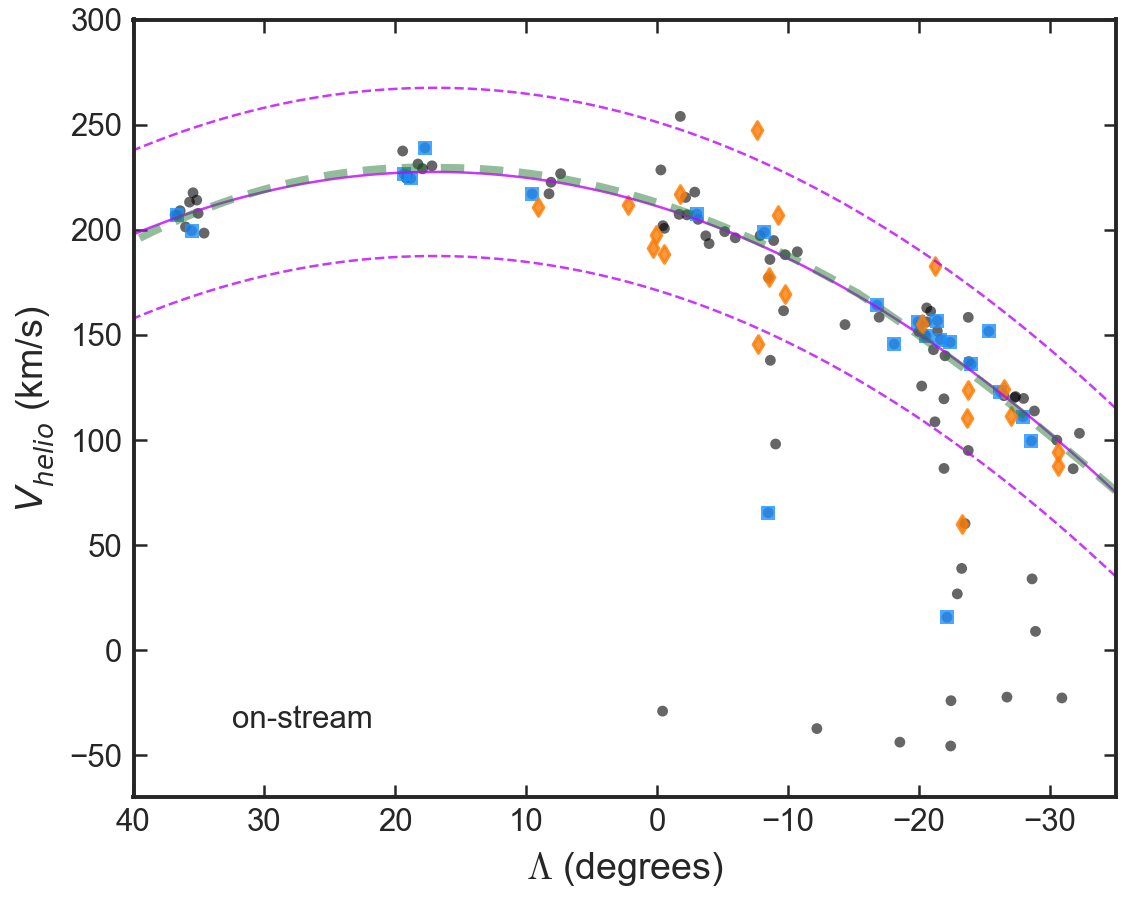}
\includegraphics[width=72mm]{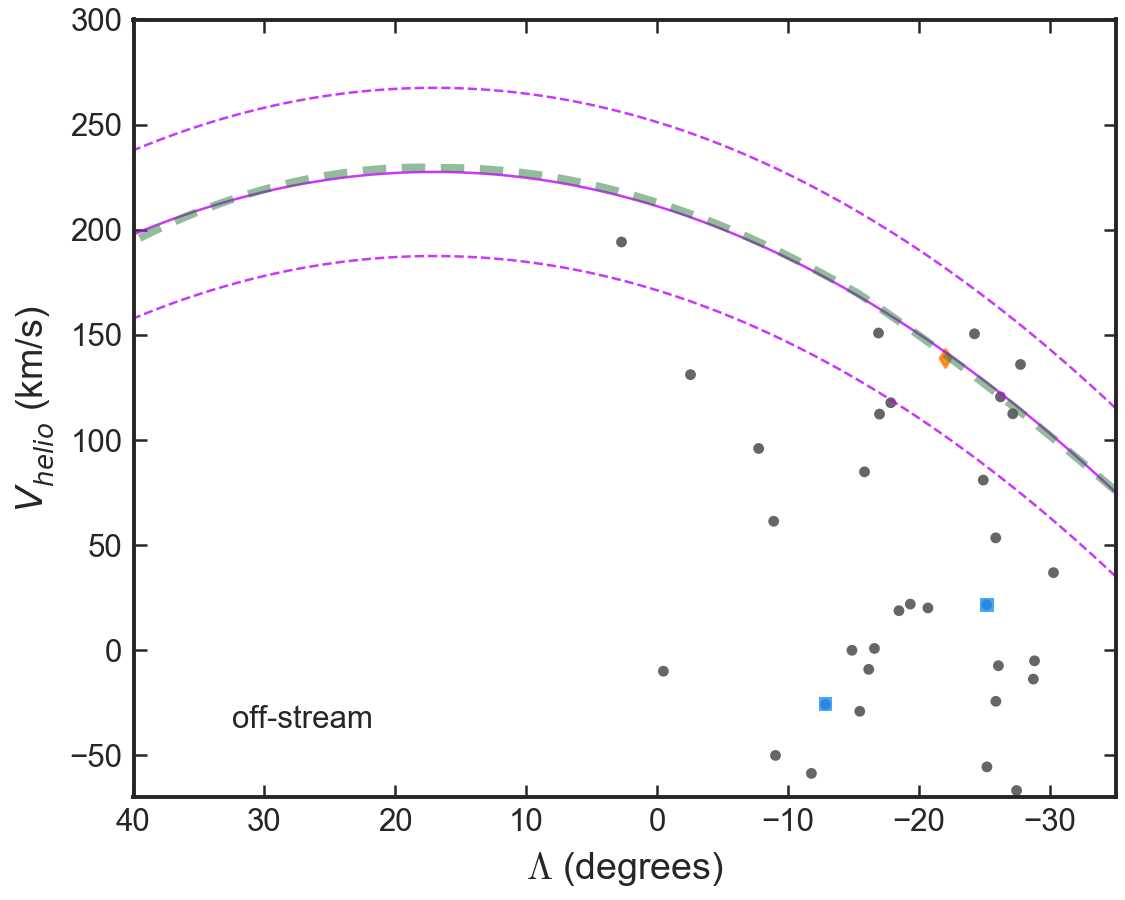}
\caption{
\label{fig.vel}
Left: velocity of stars from SDSS in the on-stream latitude region.
Black circles are selected according to the standard spatial and proper motion cuts.
Cyan squares show stars additionally passing photometric and distance modulus selection cuts
for BHB stars.
The solid line shows our fit to these points (equation~\ref{eqn.velocity}).
Dashed lines show our selection boundary of $\pm 40 \kms$ around this track.
The dashed green line shows the nearly identical track of \citet{newberg10}.
Orange diamonds show high-confidence RRL from \citet{sesar13}, with
an additional cut on proper motion made as for the other stars.
Right: same, but for the off-stream latitude region.
}
\end{figure*}

At this stage, we have useful approximations of the distance and proper motion
trends in the sample.  We would like to select RGB stars as well, to constrain
the stellar population and refine the proper motion trends.
First, though, we need to know their location
in the color-magnitude diagram (CMD).
RGB stars lie in a more contaminated region of the CMD than the BHB stars.
To obtain the cleanest possible sample, we would like to combine
parallax and proper motion cuts using \Gaia\ data with selection on 
radial velocity.

For this purpose, we obtain spectroscopic stars 
from the SDSS DR13 archive within the sky area of interest
and cross-match this sample to our previous \Gaia\ DR2 and PS1 sample.
The SDSS sample includes widely distributed stars from the ``legacy'' survey, and
spots sampled at a higher density due to the SEGUE programs.
The DR13 sample is somewhat larger than that available to \citet{newberg10}.
We continue to use PS1 as our source of photometry to enable
a homogenous selection over the largest possible sky area.
We use the fields {\tt elodiervfinal},
{\tt fehadop}, and {\tt loggadop} for the radial velocity, $\feh$,
and $\log \, g$ parameters respectively.

In Figure~\ref{fig.vel} we show the velocity trend obtained using
our standard latitude, parallax, and proper motion cuts.
The stream is easily picked out by eye, especially when comparing with
the off-stream sample.
Truncating outliers, we fit the data with the curve
\begin{equation}
v_{\mathit{helio}}(\Lambda) = (211 + 192 \, \lnorm - 563 \, \lnorm^2) \kms
\label{eqn.velocity} 
\end{equation}
We also show the track of \cite{newberg10}, translated from
their galactic standard of rest frame back to heliocentric velocity.
Despite our advantages of a cleaner and slightly larger sample,
the earlier track is almost identical.
Due to the coverage limits of SDSS, this track has only been tested over 
$-35\degree \ltrsim \Lambda \ltrsim 40\degree$,
a smaller longitude range than for the tracks in the other data dimensions.

\begin{figure*}
\subfloat[]{\includegraphics[width=44mm]{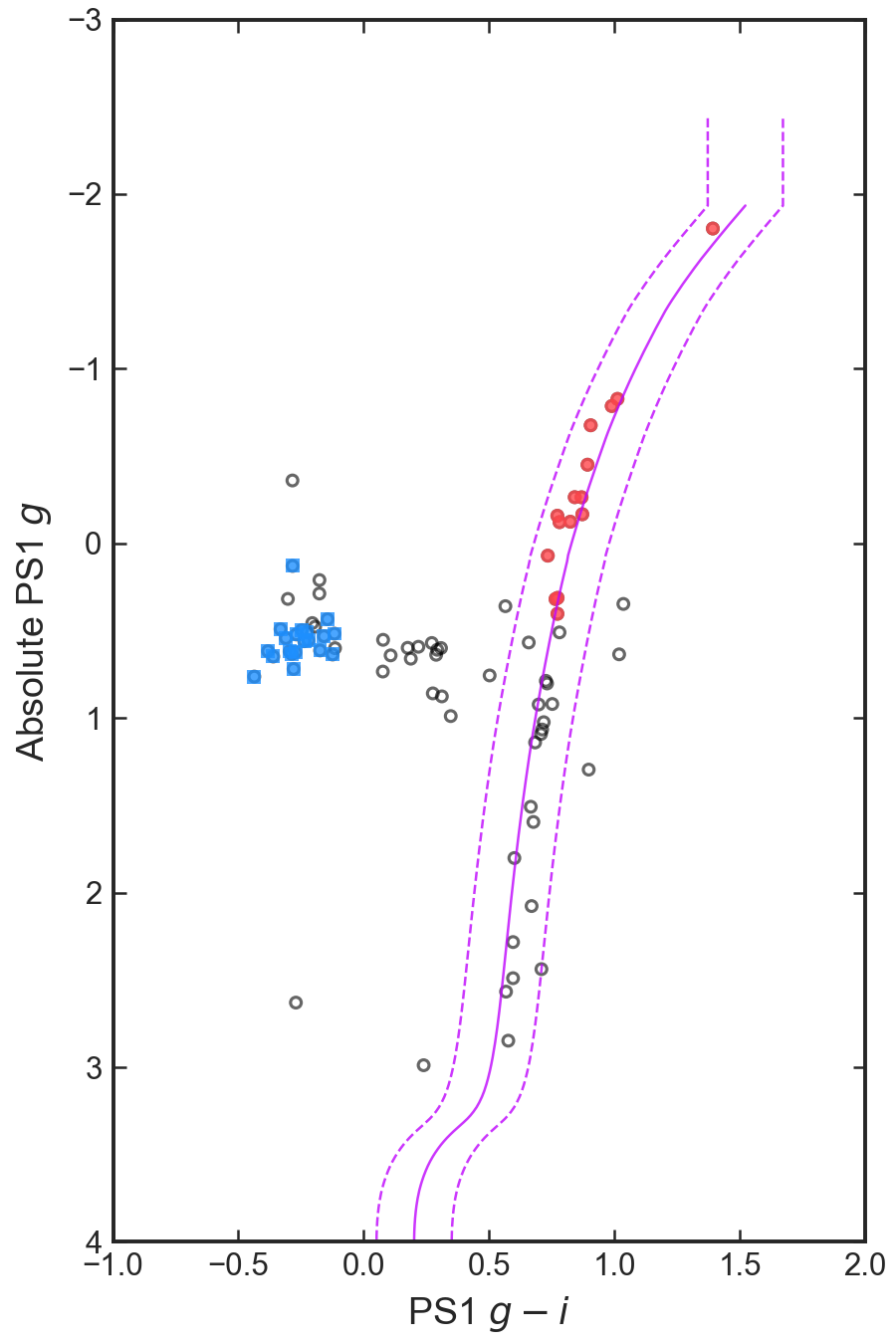}}
\subfloat[]{\includegraphics[width=44mm]{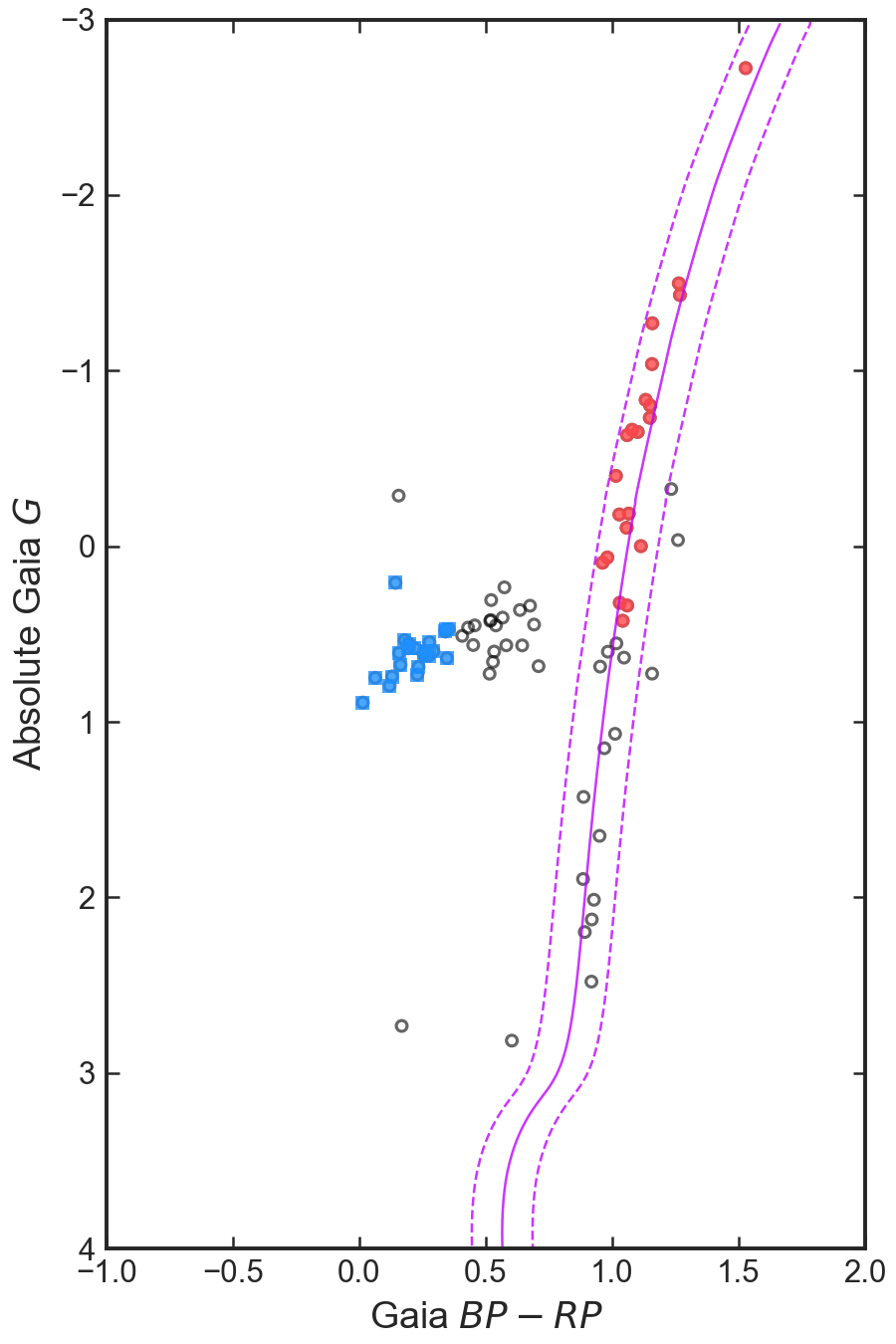}}
\subfloat[]{\includegraphics[width=44mm]{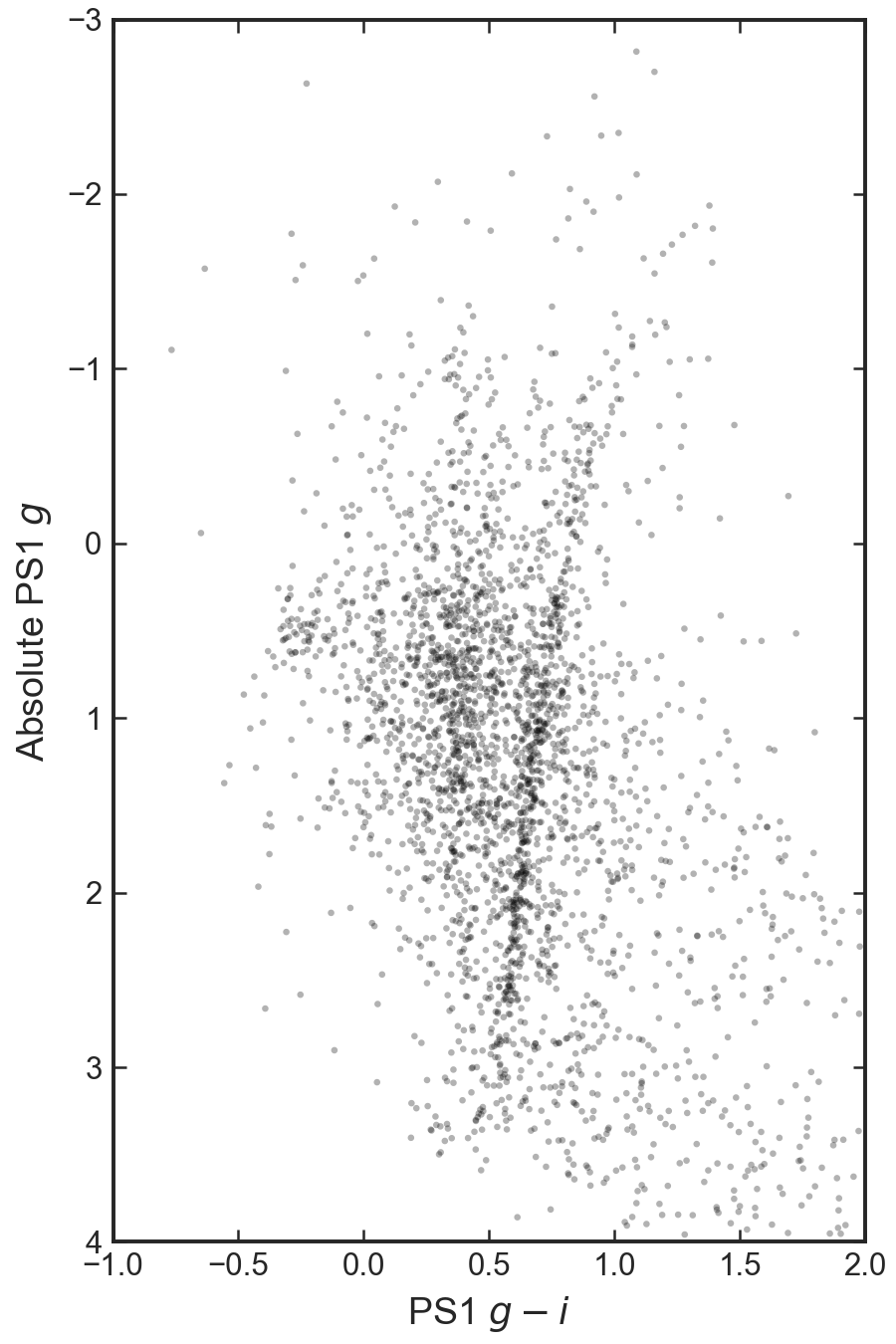}}
\subfloat[]{\includegraphics[width=44mm]{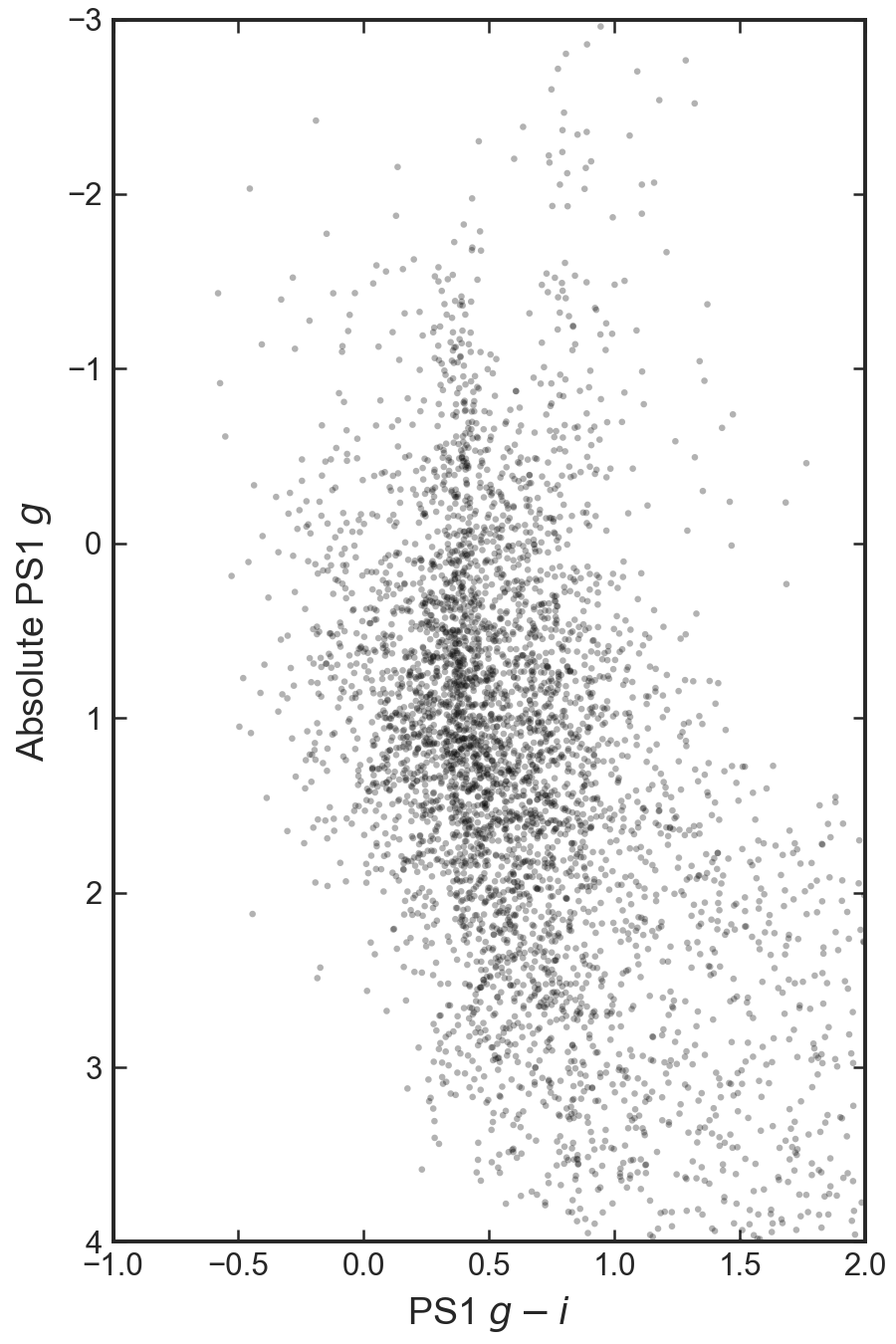}}
\caption{
\label{fig.cmd}
Panel (a): color - absolute magnitude diagram for the spectroscopically selected
sample, from PS1 photometry. 
Absolute $g$ magnitude and $g-i$ color 
are corrected for extinction and computed according to the distance modulus fit in
Equation~\ref{eqn.distancetrack}.
Points show all stars passing latitude, proper motion, and velocity cuts.  
Red points show stars passing the RGB selection cut based on PS1 photometry.
Cyan points show BHB candidates.
The solid line shows a $\feh = -1.75$, $13 \Gyr$ MIST isochrone.
Dashed lines show the RGB selection boundaries around this isochrone (we also impose
a brightness limit $M_g < 0.5$).  
Panel (b): Same, but using \Gaia\ photometry and stellar type selection criteria.
The central isochrone in this case is $\feh = -1.5$, $13 \Gyr$.
Panel (c): stars in the on-stream sky region drawn from
the overall \Gaia\ DR2~$+$~PS1 sample in the northern region.
Cuts are applied in proper motion, parallax, and data quality.
The Orphan Stream is visible as the extremely narrow
curving RGB sequence and the cloud of HB stars.  
Panel (d): same, but for the off-stream sky region.  The signatures of the Orphan
Stream seen in Panel (c) are absent.  
}
\end{figure*}

To our previous selection cuts, we now add a cut of $\pm 40 \kms$ around the
velocity track.
The stars selected in this manner are plotted in the color - absolute magnitude
diagram in Figure~\ref{fig.cmd}a.  This shows
a well-developed RGB, a strong concentration of BHB and possible RRL stars,
and even a hint of the asymptotic giant branch where it joins onto the RGB.
As noted by \citet{newberg10}, the selection of stars in SDSS is complicated
and far from uniform,
so the relative numbers of stars in different parts of the CMD (e.g.,
BHB vs RGB) are not fair reflections of the underlying population.
However, the appearance of a narrow RGB is unlikely to be an artifact.  

Using $3\sigma$ clipping,
we estimate the velocity dispersion about the mean track as $\sigma_v = 7 \kms$.  
This is close to the estimate $8$--$13 \kms$ of \citet{newberg10}.
\citet{hendel18} estimated a much larger velocity dispersion of $30 \kms$
about their orbital tracks.  As they note, however, it is difficult to
measure velocities of RRL due to their atmospheric motion, and 
the much smaller dispersions found here are probably more reliable.

We find a MIST isochrone of $\feh = -1.75$ and age $13 \Gyr$
fits the stars reasonably well.  
(We use the MIST v1.1 isochrones with $v_{\mathit{crit}}=0$ and solar $\afe$
throughout.)
From this we construct a selection boundary in the CMD (Figure~\ref{fig.cmd}a)
with width 0.2 mag on either side of the isochrone.
We also impose an absolute magnitude limit $g < 0.5$, as this keeps the
dwarf contamination low and uses only the stars with the most accurate
proper motions.
Using the stars within this CMD cut, 
the mean spectroscopic metallicity is found to be $\feh = -2.0$ with dispersion
$\sigma_{\feh} = 0.4$.
This dispersion is almost entirely real, assuming the formal uncertainties are valid, 
as the dispersion induced by observational error is $<0.2$ dex.
Given the many assumptions that go into specifying
the isochrone and the possibilities for observational error,
the offset between our photometry and spectroscopic metallicity estimates is
not particularly surprising.

In Figure~\ref{fig.cmd}b we show a similar diagram but using \Gaia\ photometry.
A distinct RGB remains, though it is slightly less tight than in
Figure~\ref{fig.cmd}a.
Here we find a better isochrone has $\feh = -1.5$ with age $13 \Gyr$.
We use a selection boundary with half-width 0.15 mag to choose stream stars.
Again, we will not be troubled by the slight disagreement with the
spectroscopic metallicity.

The stellar population of the stream is also vividly illustrated using the
overall \Gaia\ PS1 sample without any spectroscopic selection,
with much higher signal though also higher contamination.
The stream signature is obvious in the on-stream region in Figure~\ref{fig.cmd}c,
whereas it is absent in the off-stream region in Figure~\ref{fig.cmd}d.
Although we have omitted the isochrone to improve the plot clarity, it agrees with
the narrow RGB down to faint magnitudes, well below our absolute magnitude cut.
The RGB is slightly wider but still easily visible when using \Gaia\ photometry alone.

\citet{casey13} obtained spectra of stream candidates in a field spanning the
longitude range $15$--$25 \degree$,
and identified 9 stars as stream members.
It turns out that while 5 of them are in
the proper motion range we identified as belonging to the stream,
4 of them are not.
Of the 4 non-members (OSS 4, 9, 12, and 19),
3 have higher metallicities than the rest of the sample.
As with the \citet{sesar13} sample, 
pruning the sample with proper motion
reduces the mean metallicity and dispersion in this sample,
but does not eliminate the dispersion.
The revised mean metallicity from the 5 remaining stars
would be $-2.1$, with dispersion 0.5.  
\citet{casey14} obtained high-resolution spectra and improved metallicity
estimates of 3 of the \citet{casey13} stream candidates (OSS 6, 8, and 14).
All three of these pass our proper motion cut
and are thus highly likely to be stream members.
These stars span a range of over 1 dex in $\feh$, supporting
a substantial metallicity dispersion within the stream and
disfavoring a globular cluster as the progenitor.
\subsection{Proper motion}
\label{sec.pm}
\begin{figure}
\onecol{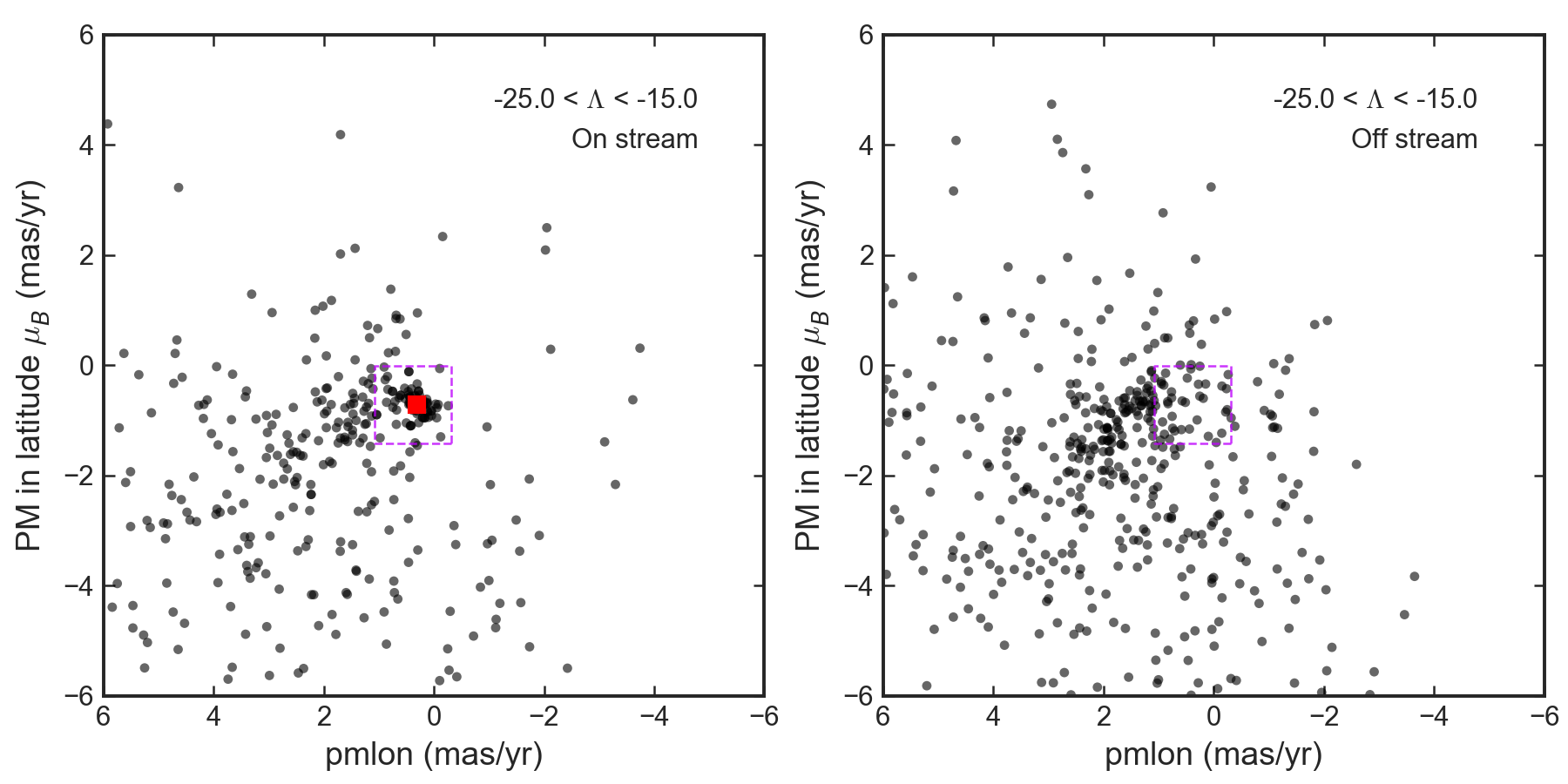}
\onecol{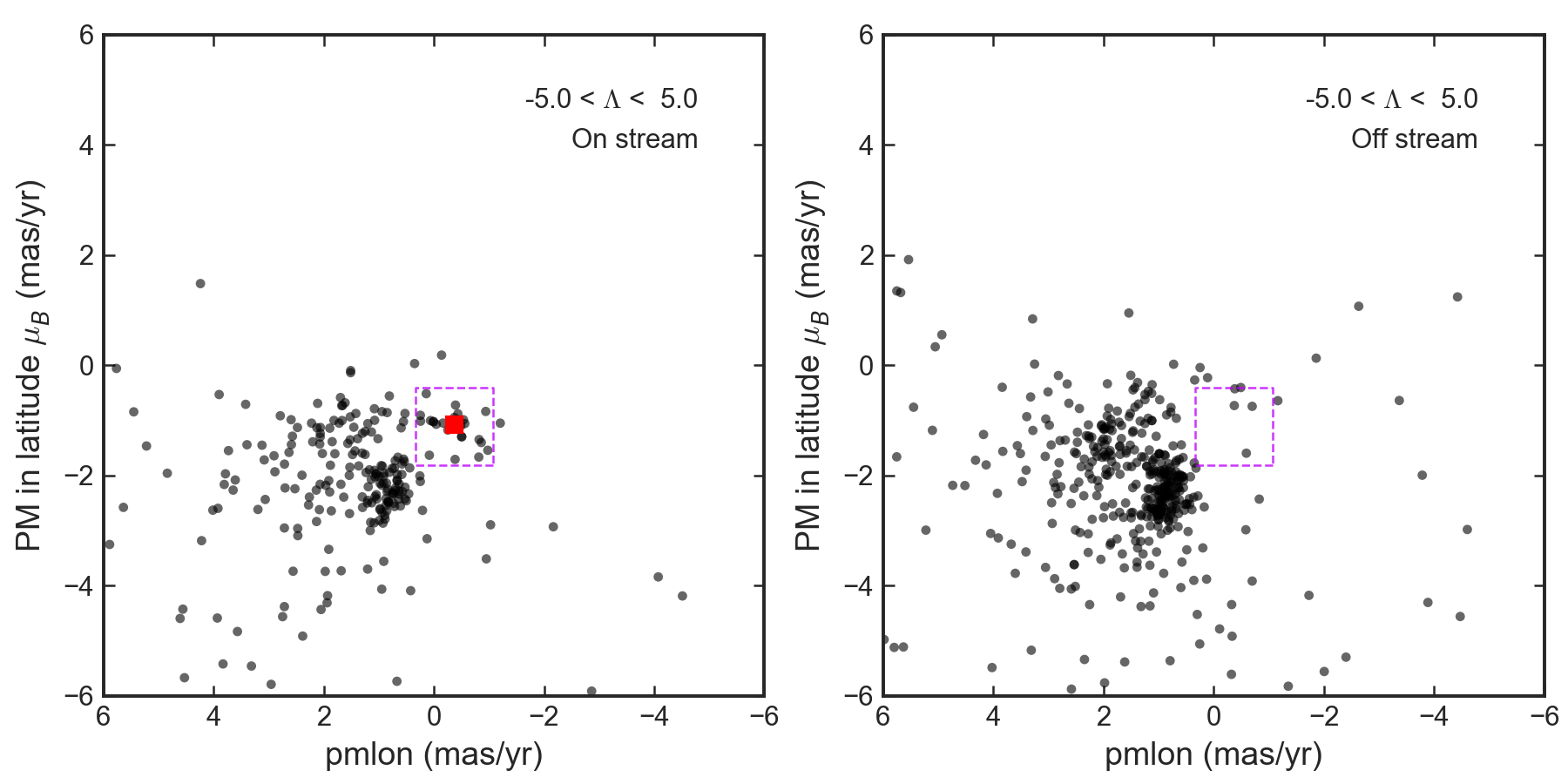}
\onecol{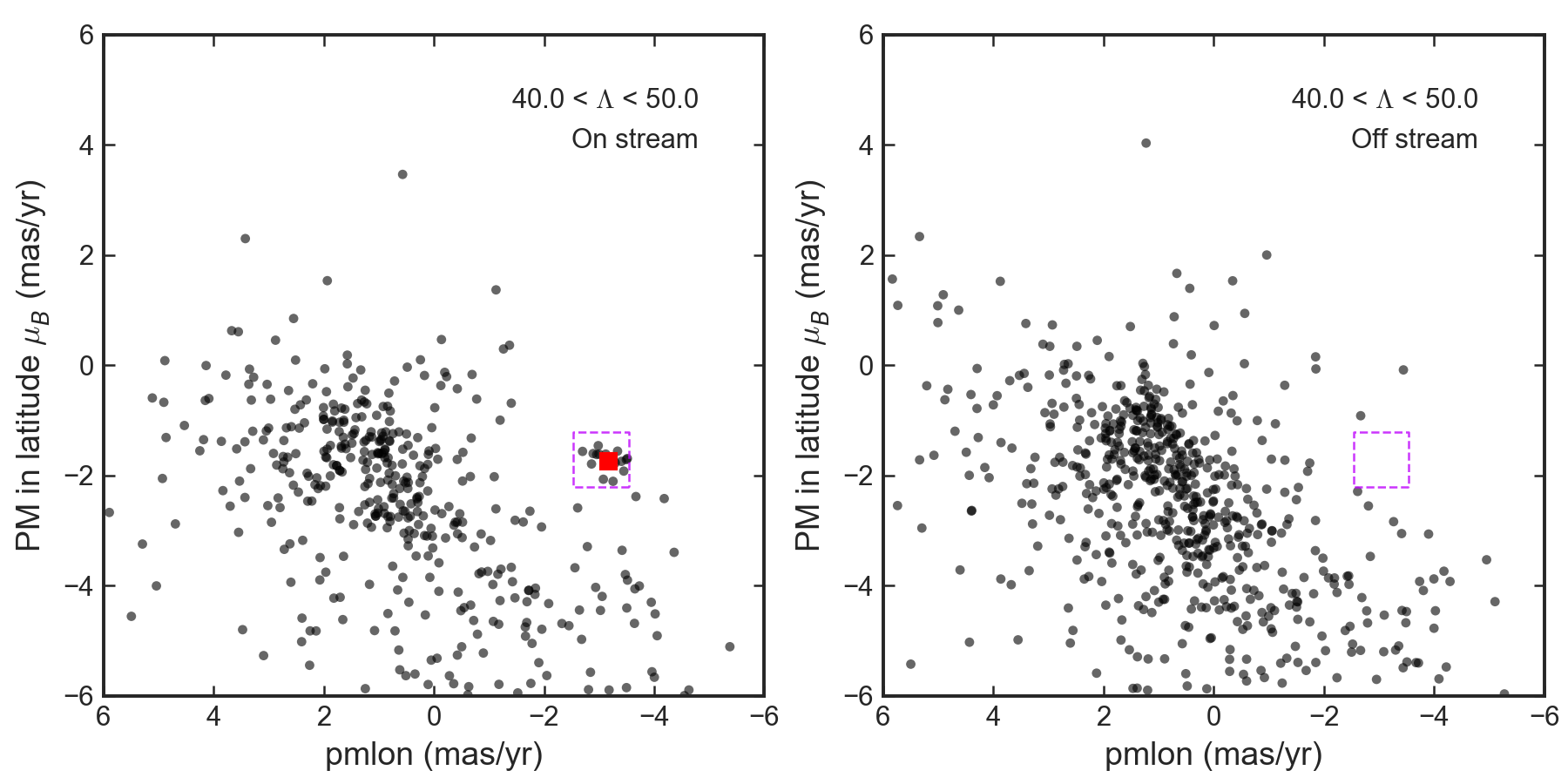}
\caption{
\label{fig.pmspace}
Proper motion of stars in the Orphan Stream region.
Each row shows a different bin of $10\degree$ in longitude $\Lambda$.
Left column: on-stream region defined by a latitude interval
centered on the stream track.
Right column: off-stream region, where the stream concentration is absent.
Black points show sample stars, including RGB, RRL, and BHB stars,
selected based on photometry as appropriate to stellar type.
The red point shows the fit to the mean stream proper motion in this bin.
The dashed box shows the square PM selection box,
which is centered on the smooth track of Equation~\ref{eqn.pm}.
The dense clump near $(-1 \masyr, -2 \masyr)$ in both panels of the second row is
from the Sagittarius Stream (see text).  
}
\end{figure}

\begin{figure*}
\twocol{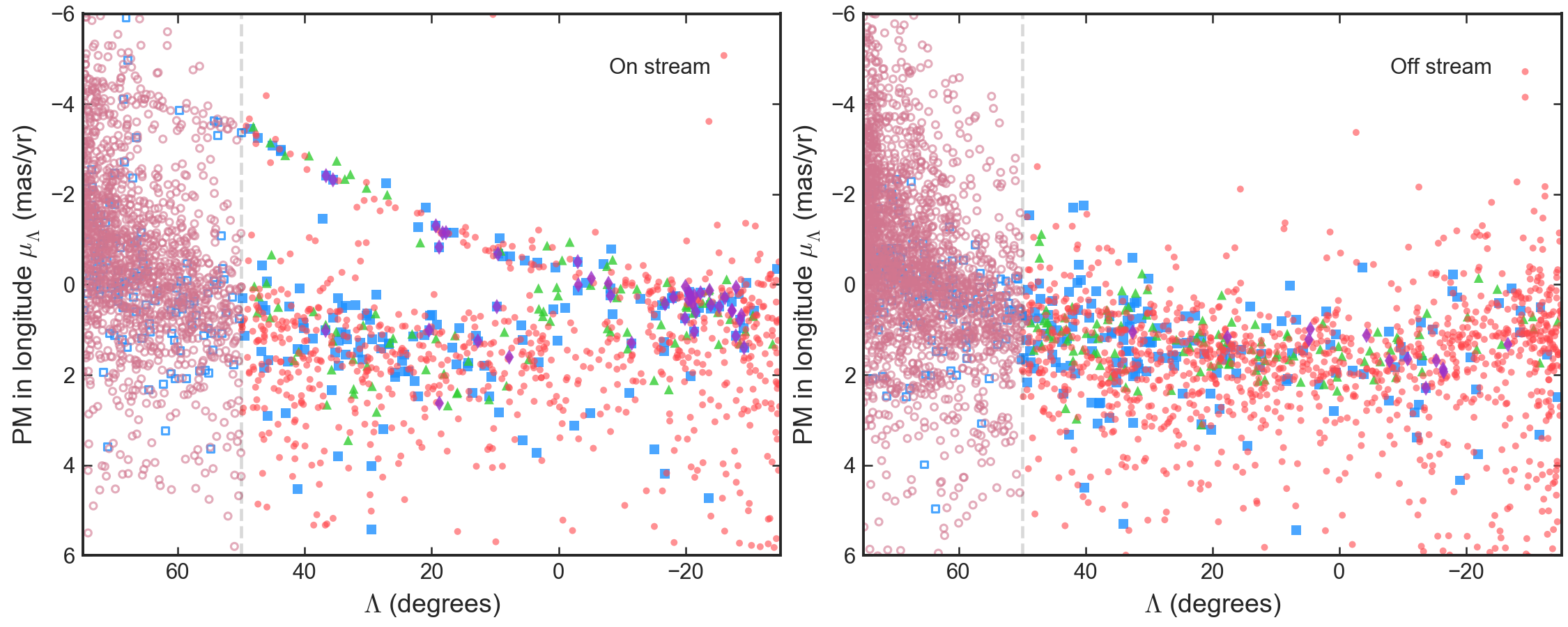}
\twocol{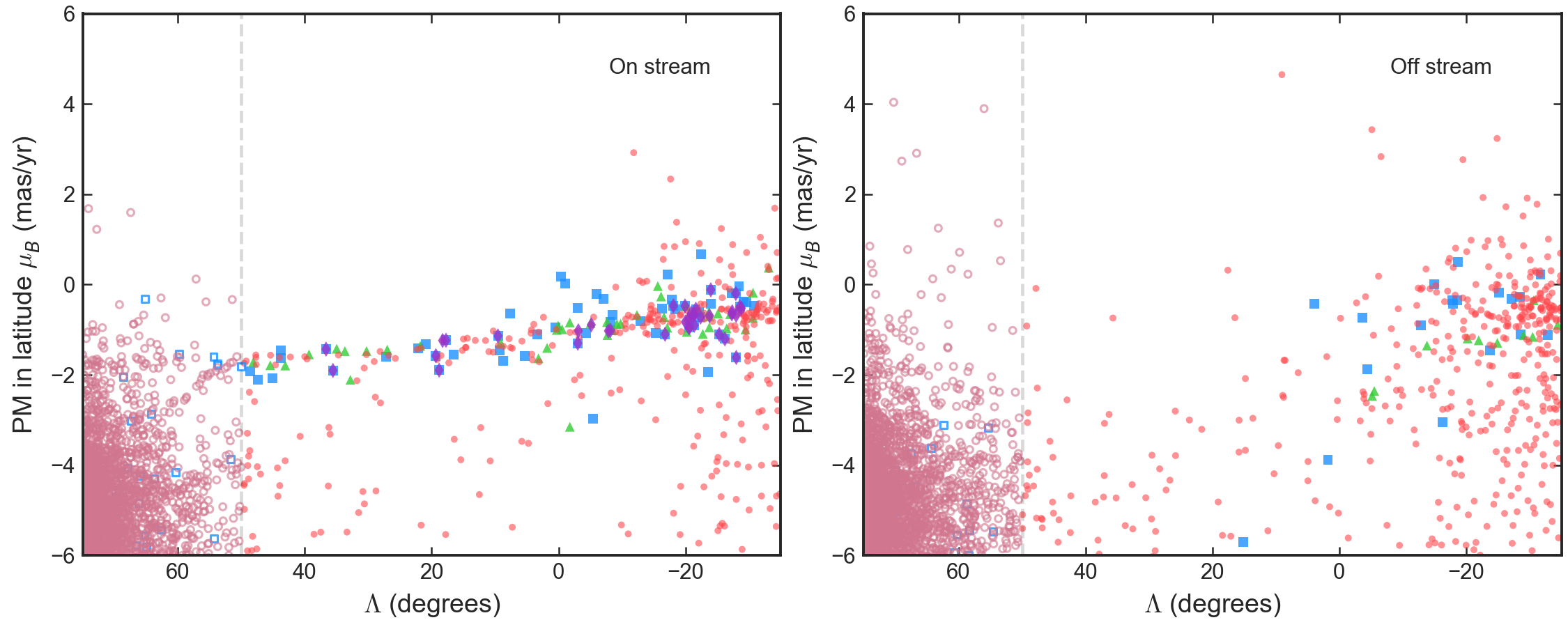}
\caption{
\label{fig.pmstars}
Proper motions of stars along the Orphan Stream path.
Top and bottom rows show proper motions in longitude and latitude $(\pmlon, \pmlat)$.
Left column: on-stream sky region selected by a latitude cut around the stream track.
Right column: off-stream sky region.
Data points show individual stars coded according to selection method.
Solid cyan squares: northern BHB;
empty cyan squares: southern BHB; 
green triangles: RRL;
solid red circles: northern RGB;
empty pink circles: southern RGB;
purple diamonds: velocity-selected spectroscopic sample.
For each dimension, cuts are applied to the stars in all other observable dimensions---i.e. for the
$\pmlon$ plots, stars are selected based on $\pmlat$, and vice versa.
All panels also use selection on spatial position (on or off stream),
and distance modulus or CMD cuts as appropriate to the stellar type.
The Orphan Stream is visible as the narrow band following the high-purity spectroscopic points.
}
\end{figure*}

We already used an initial proper motion cut based on the BHB and RRL subsamples
to help select stream stars and define the behavior in other dimensions.
Now we add the information from RGB stars, some of which are quite bright and
therefore have small proper motion uncertainties, to define the proper
motion trend more precisely.

We divide our entire sky region into longitude bins $10 \degree$ long,
overlapping by $5 \degree$ so that
only every second bin is independent.  We combine our BHB, RRL, and RGB
samples selected by parallax and by distance modulus (for BHB and RRL)
or color-magnitude position (for RGB) as described above.
We switch between the northern and southern selection methods at $\Lambda = 50 \degree$.
In each bin, we select stars inside and outside our latitude cut of $\pm 2.25\degree$ around the
stream track for the signal and background samples respectively.
We use only stars within the box $|\pmlon| < 6 \masyr$, $|\pmlat| < 6 \masyr$.
This separates our estimation of the stream proper motion from details of
the distribution at much higher proper motions, which correspond to nearby disk stars.

The combined sample in several bins is shown in Figure~\ref{fig.pmspace}.  In each bin,
a clump of stars associated with the stream is apparent,
though in some bins it is stronger or more distinct than in others.
Over a latitude range $-20\degree < \Lambda < 10\degree$, a strong
second cold clump is apparent in both the on-stream and off-stream samples.
From its sky position (RA $\approx 153\degree$, declination $\approx 23\degree$)
and proper motion, this clump is
identifiable as the leading arm of the Sagittarius Stream.

\begin{figure*}
\twocol{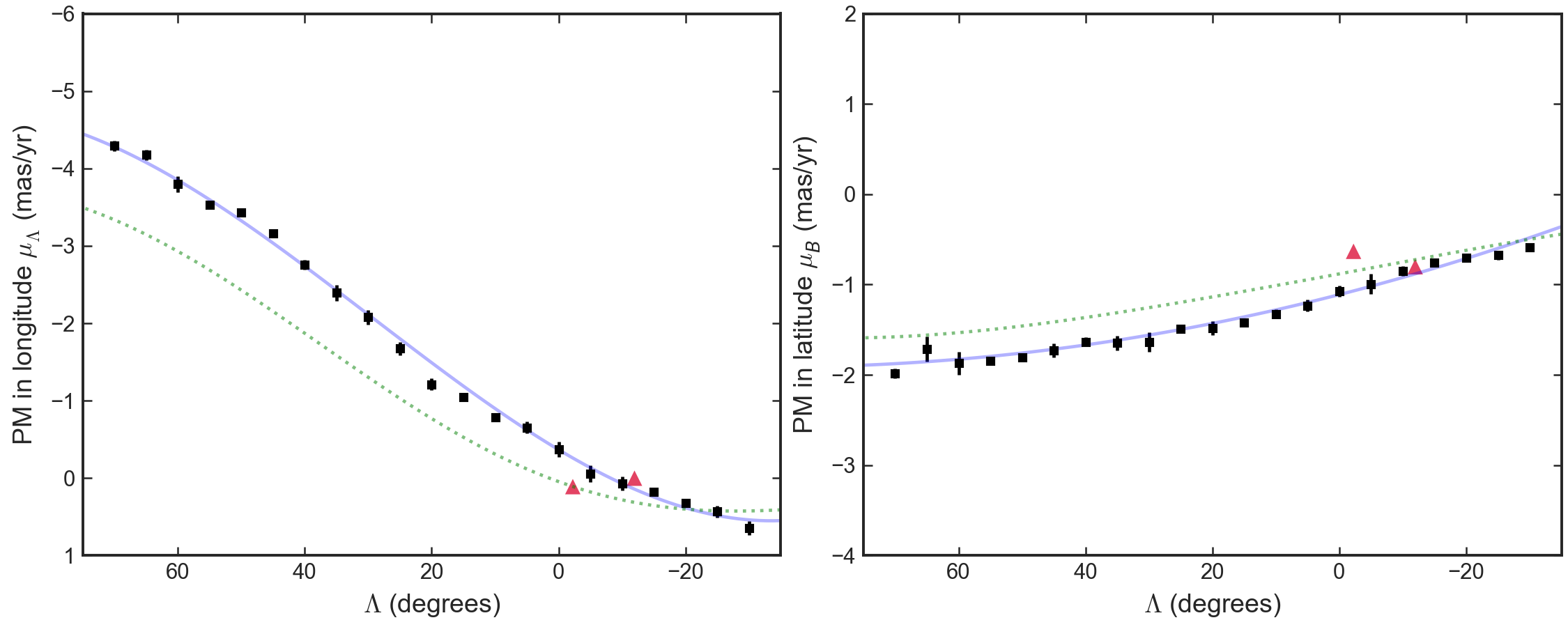}
\caption{
\label{fig.pmtracks}
Proper motion tracks of the Orphan Stream in longitude (left) and latitude (right) directions.
The data points show results from fits to $10\degree$ bins.
The bins overlap by $5 \degree$ so only every second bin is independent.
The solid line shows our fit to these points (equation~\ref{eqn.pm}).
Red triangles show the detections of \citet{sohn16}.  
The dotted line shows orbit~5 from \citet{newberg10}.
}
\end{figure*}

We first use the Gaussian Mixture Model code
{\tt \href{https://github.com/pmelchior/pygmmis}{pyggmis}}
to fit the background (off-stream) sample of each bin
in the space of $(\pmlon, \pmlat)$, using a Gaussian mixture of either
2 or 3 components depending on the apparent complexity of the off-stream sample
in that bin.  
We choose this particular code because it
corrects for the censored data outside our proper motion box.
We then add another component to represent the stream in the signal region,
and initialize the guess for this component's center to the value 
given by our previously derived trend.
We initialize the other components to the values given by the background fit.
We run {\tt pyggmis} on the signal region data to obtain a fit to the stream proper motion.
We then bootstrap-resample the points and repeat the procedure
to obtain estimates of the uncertainties.
The final fitted values are shown by the points in Figure~\ref{fig.pmspace} centered on the stream clump.

Once we have a fit to the mean stream proper motion within each longitude bin,
we fit the overall trend with
a cubic in $\Lambda$ for $\pmlon$ and a quadratic for $\pmlat$, yielding the tracks
\begin{gather}
\pmlon = (-0.31 - 4.73 \,\lnorm - 6.04 \, \lnorm^2 + 6.86 \, \lnorm^3) \masyr    \notag \\
\pmlat = (-1.10 - 1.85 \, \lnorm  + 1.03 \lnorm^2) \masyr
\label{eqn.pm}
\end{gather}
The trends of the on and off-stream samples with latitude are shown with
individual stars in Figure~\ref{fig.pmstars}.
Here we use Equation~\ref{eqn.pm} to select stars in the proper motion dimension that is {\it not} plotted,
along with our other usual sample cuts.
(Without this additional cut, the contrast of the stream versus the background would
be greatly reduced.)
Left panels show the on-stream sky region, and
right panels the off-stream region.  It is easy to pick out the stream in the on-stream regions,
while it is essentially absent from the off-stream region.
Some interesting hints of substructure are present through the increased dispersion at
certain locations and the possible kink in $\pmlat$ near $\Lambda = 50\degree$.
We will not pursue these further here.

The individual bin fits and the track given by Equation~\ref{eqn.pm} are displayed
in Figure~\ref{fig.pmtracks}.  
The PM estimates of \citet{sohn16} are shown by the two red triangles.
These estimates were obtained by finding stars roughly matching the 
main sequence at the distance of the stream as well as the proper motion of the
stream as predicted by \citet{newberg10}.
Converting the \citet{sohn16} values to stream coordinates, the field at
$\Lambda = -12.0\degree$ has $\pmlon = 0.0 \masyr$, $\pmlat = -0.80 \masyr$,
and the field at 
$\Lambda = -2.2\degree$ has $\pmlon = 0.11 \masyr$, $\pmlat = -0.63 \masyr$. 
The point at $\Lambda = -12 \degree$ is in excellent agreement with our results.
The point at $\Lambda = -2 \degree$ deviates somewhat in both dimensions, which is not
surprising---this point represents the lone star in that field
consistent with expected stream properties.

Figure~\ref{fig.pmtracks} include the prediction of orbit~5 from \citet{newberg10} (their best fit),
which we recomputed with the aid of the
{\tt \href{https://github.com/jobovy/galpy}{galpy}} package.
This orbit agrees reasonably well with our fit
around $\Lambda=-20\degree$.  Its slope and curvature also agrees qualitatively with
our results in both dimensions.
Quantitatively, however, this orbit is ruled out by our results at very high significance, 
and the absolute differences reach as high as $\gtrsim 1 \masyr$.

To define sample cuts based on proper motion, we center our
selection box on the fit given by Equation~\ref{eqn.pm}.
In the northern region we use a box of a fixed half-width $0.7 \masyr$ in
each proper motion dimension $(\pmlon, \pmlat)$ separately,
centered on the trend of Equation~\ref{eqn.pm}.
In the southern region the proper motion errors are smaller
and the contamination of the RGB and BHB samples is greater, 
so we decrease the box half-width to $0.5 \masyr$.

Since stream coordinates for the proper motion may not be preferable in all cases, 
we re-express the proper motion trends in Equation~\ref{eqn.pm} in equatorial coordinates
with the approximate fits
\begin{gather}
\mu_{\mathit{RA}}  = (-1.17 - 3.33 \, \lnorm - 0.13 \, \lnorm^2 - 0.22 \, \lnorm^3) \masyr    \notag \\
\mu_{\mathit{Dec}} = (-0.06 + 4.21 \, \lnorm + 4.89 \, \lnorm^2 - 6.88 \, \lnorm^3) \masyr
\label{eqn.pmradec}
\end{gather}
In galactic coordinates we find
\begin{gather}
\mu_{\mathit{l}} = (-0.42 - 7.35 \, \lnorm -  4.63 \, \lnorm^2 + 19.72 \, \lnorm^3 \notag \\
\hspace{4em} - 12.58 \, \lnorm^4) \masyr    \notag \\
\mu_{\mathit{b}} = (-1.04 - 0.27 \, \lnorm + 11.64 \, \lnorm^2 \notag +  2.12 \, \lnorm^3 \notag \\
\hspace{4em} - 12.37 \, \lnorm^4) \masyr
\label{eqn.pmlb}
\end{gather}
The accuracy of these fits should be better than $0.1 \masyr$,
except possibly near the ends of the
observed range $-35\degree < \Lambda < 75\degree$.
\subsection{Sky position}
\label{sec.sky}
\begin{figure*}
\twocol{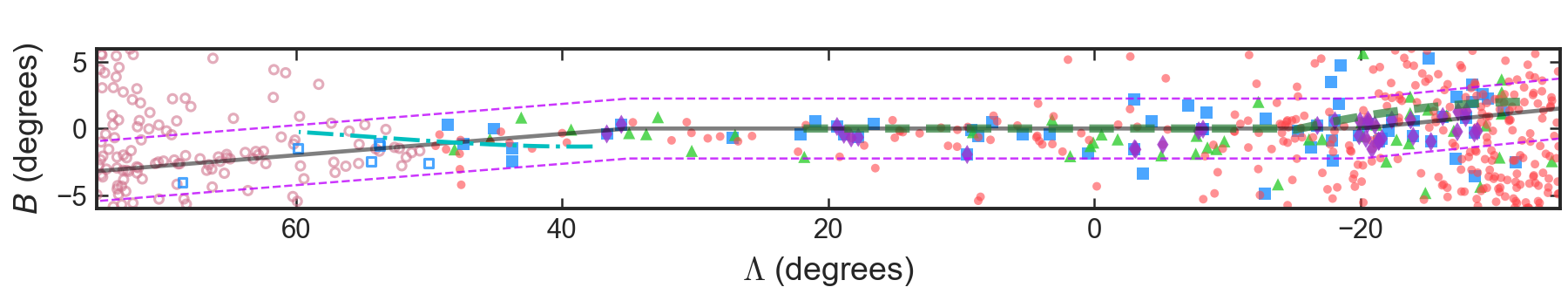}
\caption{
\label{fig.sky}
Spatial distribution of sample sources.  Point types correspond to stellar type
as in Figure~\ref{fig.pmstars}.
Stars are selected according to photometry, proper motion, 
and (for the spectroscopic sample) velocity.  
The stream is quite clean and concentrated in $0\degree \ltrsim \Lambda \ltrsim 50\degree$ or so,
while the contamination increases at both low and high $\Lambda$.
Nevertheless a concentration is clearly visible in the north (low $\Lambda$)
roughly following the bend in the track found by \citet{newberg10} (green dashed line).
Cyan line shows the track of \citet{grillmair15}.
In the south (high $\Lambda$), the stream appears to veer towards negative $B$
at least up to $\Lambda = 65\degree$ and perhaps further.
Solid black line shows our adopted stream track.
Dashed magenta lines show our on-stream latitude selection cut, $2.25\degree$
on either side of this track.
}
\end{figure*}

To refine the track of the stream on the sky,
we now select stars according to the parallax, proper motion, and
distance modulus or color-magnitude properties of the stream.
Figure~\ref{fig.sky} shows the stars from our various samples plotted
on the sky in stream coordinates.
Here increasing contamination is apparent
in the north (low $\Lambda$)
because the proper motion cut loses discriminating power there, due to the smaller
separation from the background stars.  
Still, it appears the stream roughly follows the bend to
larger latitude $B$ at low $\Lambda$ already found by \citet{newberg10}.

In the south (large $\Lambda$), the increasing contamination again makes the
stream difficult to follow.  Nevertheless, we find
the stream stars have their peak density
at lower and lower $B$ as $\Lambda$ increases, reaching a deviation of at least
$2 \degree$ from the equator of the coordinate system.
The Orphan Stream in the south was previously mapped by \citet{grillmair15}
using stars near the main-sequence turnoff.  They found a change
from a fairly well-defined stream over $-38 \degree < \delta < -18 \degree$
($38 \degree < \Lambda < 60 \degree$) to a broader and bifurcated
structure further south.  They attributed the brightest structures in
the southern section to inaccurate correction of the highly structured extinction
in this area.  
Our stream map consists of stars from entirely different parts of the
color-magnitude diagram than in \citet{grillmair15}, yet we find an
overdensity in the same regions as the brightest overdensities in
their maps.  This suggests these overdensities may be real.
Our stream path does not closely follow the analytic fit provided by
\citet{grillmair15} or show the S-shaped bend at $\delta=-14\degree$
($\Lambda=34\degree$), but the density of our tracers is particularly
low in this area so it is not clear if there is a real disagreement.

Given the hints of irregular spatial structure and the relatively small signal,
we have not performed a fully automatic fit
to the stream track.  Instead, we matched a piecewise linear trend,
modifying the earlier trend of \citet{newberg10},
to follow the apparent stream overdensity in Figure~\ref{fig.sky}:
\begin{gather}
B = -0.10 \, (\Lambda + 20\degree), 
\; -35\degree < \Lambda < -20\degree   \notag \\
B = 0, 
\; -20\degree < \Lambda < 35\degree   \notag \\
B = -0.08 \, (\Lambda - 35\degree), 
\; 35\degree < \Lambda < 75\degree  \\
\end{gather}
We use an interval of $2.25 \degree$ around this track when selecting stars for fitting
or displaying in the other dimensions.  In the central range $0\degree < \Lambda < 50\degree$
where the stream is best defined, we estimate a dispersion of approximately $\sigma_B = 1.3\degree$
around the track.  This agrees reasonably well with the estimated dispersion of
$0.9 \degree$ from main-sequence stars in \citet{belokurov06}.

As a final check, we examined images of the overall sample of stars
selected according to our various cuts
for those lying within the
SDSS or PS1 surveys.  The vast majority appear to be ordinary single
stars.  A very few might have their measurements affected by nearby
stellar or galactic sources, but these are so rare that any effects on
our results are insignificant.

\section{INTERPRETATION}
\label{sec.interpretation}
\begin{figure*}
\twocol{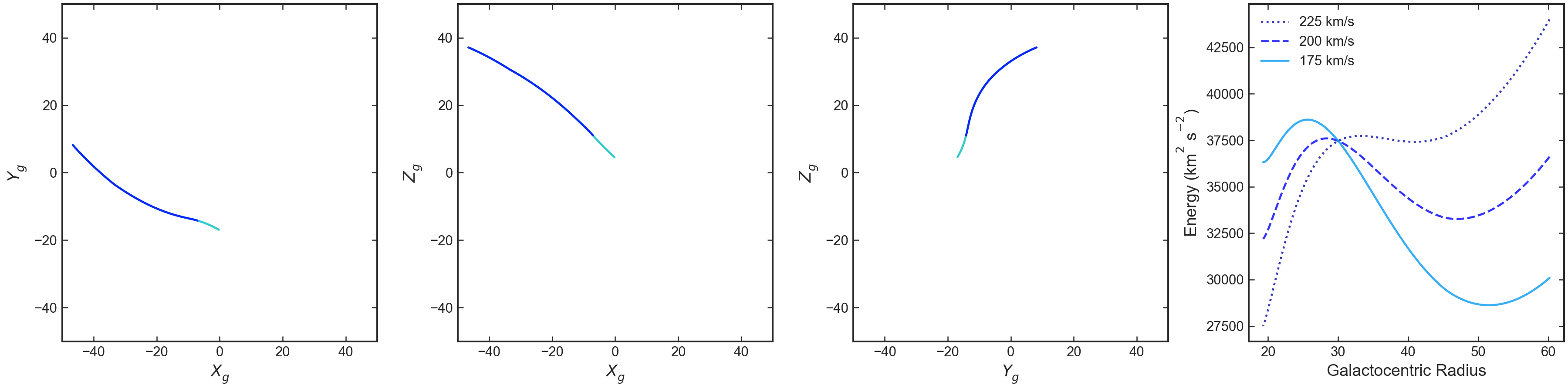}
\caption{
\label{fig.gcen}
Left three panels: track of the stream in Galactocentric coordinates.
The darker blue line shows the portion of the track $-35\degree < \Lambda < 40\degree$
where all 6 dimensions are constrained, while the short, light blue continuation
$40\degree < \Lambda < 70\degree$ is the interval lacking radial
velocity constraints.
Here the Sun is at $X_g=8.12 \kpc$, $Y_g=Z_g=0$.
Right panel: total energy of the stream as a function
of Galactocentric radius, calculated using the six-dimensional track, and
assuming logarithmic halo potentials with three different circular velocities.
The potential is normalized to zero at $r=30 \kpc$.
}
\end{figure*}
The 6d track we have obtained enables us to examine the physical 
properties of the Orphan Stream.  We keep this investigation brief since
the near future will probably bring further significant information on the stream.
It is not clear the full extent in longitude of the stream 
has been detected, and follow-up spectroscopy of likely stream
targets should specify the velocity track over a greater longitude range.

In Figure~\ref{fig.gcen}, we plot our empirical
track of the Orphan Stream in Galactocentric coordinates.
Here we assume a distance to the Galactic center of
$8.12 \kpc$ \citep{gravity18}.
We define $X_g$ so that it increases from the Sun to the Galactic center,
in the same direction as Galactic $X$.
The currently detected portion of the stream is a curving track over $60 \kpc$ long
which approaches the disk plane at its
southern extremity.  A twist in this closest portion is apparent in these
panels, though this part of the track is particularly uncertain due
to sparse distance tracers and heavy contamination from the disk.

A tidal stream consists of stars moving with similar though not exactly constant
energies.  
In the last panel of Figure~\ref{fig.gcen}, we plot the energy of the stream 
versus Galactocentric radius, assuming logarithmic halo potentials of three 
different circular velocities.  The zero-point of the potential
is set at $30 \kpc$.
We see that the variance of the inferred energy is minimized at about
$200 \kms$, suggesting an outer Galactic rotation curve of about that level.
While this is only a crude test,
this value is in good agreement with various models of Galactic potential
\citep{blandhawthorn16},
and in particular with recent results using \Gaia\ data on various tracers
\citep{watkins18,posti18,wegg18,vasiliev18}.  
We note that there is still a substantial wiggle in the energy versus radius,
and it seems unlikely we could choose a simple potential where the
energy could be made constant.

We fit several orbits to the observed stream track, using a fixed Galactic potential.
The main goal here is not parameter fitting, 
because we know that streams have systematic departures
from orbits \citep{johnston98, eyre11},
and thus we do not aim for statistical rigor.
Rather, we want to look for systematic differences between the stream and
the fitted orbits.
We fitted our orbits to our derived stream tracks at a set of points
spaced equally in longitude, rather than directly fitting the data.
We assigned rough uncertainties of $\tsim 1/2$ the selection window
in each observed dimension to govern the fits.

Three example orbits are listed in Table~\ref{tab.orbits}.
In orbit 1 we combine the solar position of $8.12 \kpc$ with
the proper motion of the Galactic center
$6.379 \masyr$
\citep{reid04},
and the motion of the Sun relative to the local standard of rest
$U = 11.10 \kms$, $V = 12.24 \kms$, and $W = 7.25 \kms$ \citep{schoenrich10}.
The potential consists of a Hernquist bulge, a Miyamoto-Nagai disk, and an
Navarro-Frenk-White halo, as specified in the table caption.

\begin{figure}
\onecol{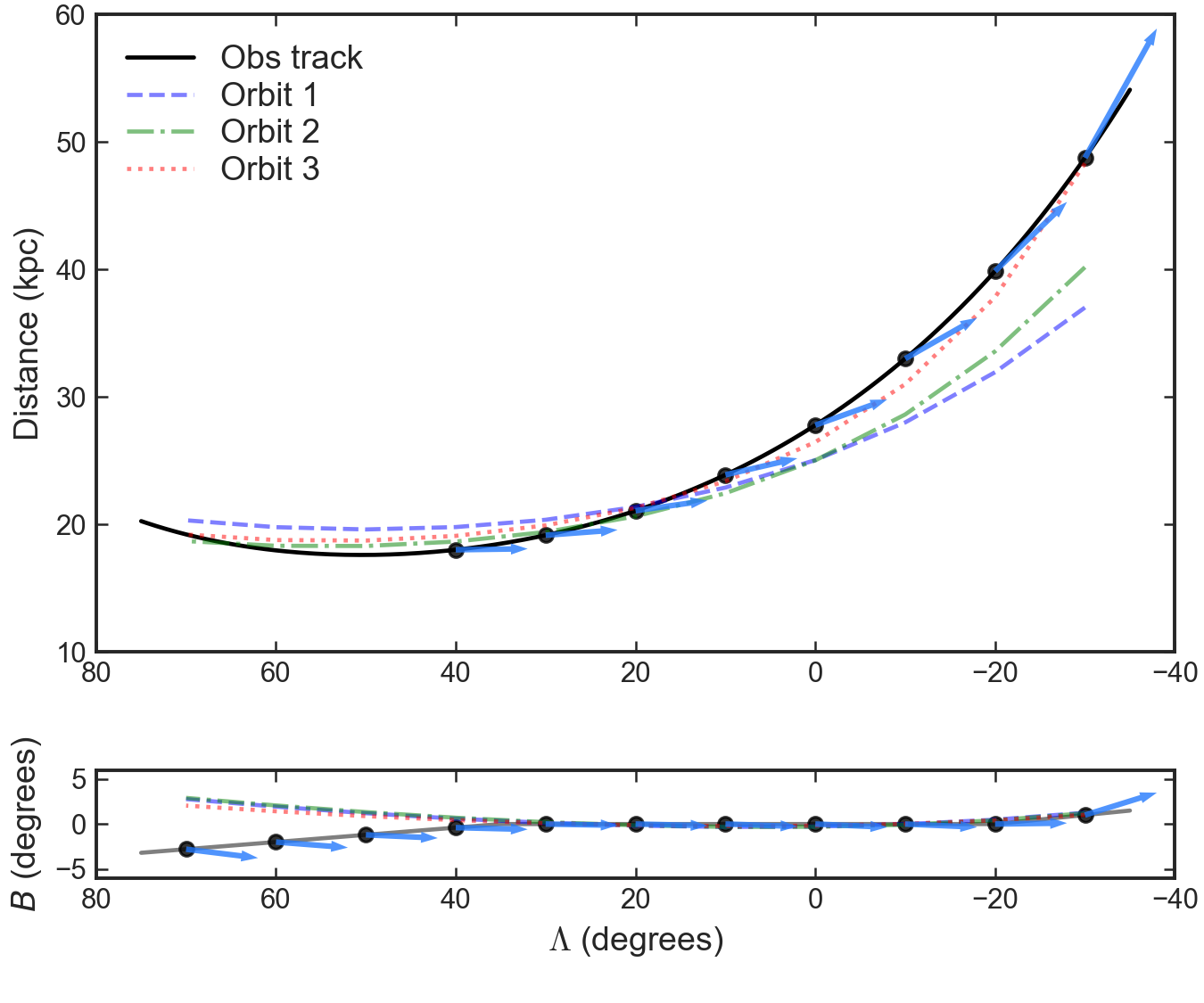}
\caption{
\label{fig.misalign}
Comparison of the stream spatial track (black solid line) and
local velocity vector (blue arrows).
The three orbits in Table~\ref{tab.orbits} are also shown.
The unit galactocentric velocity vector is shown arbitrarily scaled
for ease of visualization.
Top: Distance versus longitude.  The longitude range of the arrows
is restricted to the interval
probed by SDSS spectroscopy.  A significant misalignment is seen, in
that the velocity points less radially
than the spatial track.  Bottom: latitude versus longitude.  The velocity
arrows span the full latitude range of our track, since in this
projection the radial velocity is not needed to compute their
direction.  Misalignment is mainly visible at large $\Lambda$ where
the stream track bends to negative $B$.  In both panels the velocity
is computed using our smooth empirical tracks, and transformed to the
galactocentric reference frame using our standard solar motion
parameters (see text).  If the stream
track were an orbit, the arrows would point along the track.}
\end{figure}

Orbit~1 unsurprisingly comes much closer to matching the
proper motion than the orbit 5 of \citet{newberg10}, though some discrepancies remain.
However, it is unable to fit the derived track in several respects.
The most significant is that the distance does not reach values as large
as in the observed stream, and the radial velocity is too high.
The reason is simple: {\it the spatial path of our derived stream
track does not point in the same direction as its three-dimensional velocity vector,
so it clearly cannot be modeled well by any orbit.}
The misalignment between the spatial path and velocity vector
reaches approximately $20\degree$ in places.
Figure~\ref{fig.misalign} shows two projections, with the
solid line representing our track and the arrows the
direction of the local stream velocity after
correcting for the solar reflex motion.
As shown in the top panel, the stream path moves outward more rapidly
along the direction of the stream's motion (with $\Lambda$ decreasing)
than one would expect from the velocity vector.
In the bottom panel, the velocity does not point along our inferred
path of the stream at large $\Lambda$ (the southern part of the stream).
Our orbit fits also have difficulty in following the S-shaped path
in latitude, and tend steadily towards larger $B$ at large $\Lambda$,
rather than following the bend to negative values in our empirical track,
as shown in the figure.

This misalignment was already somewhat apparent in the work of
\citet{hendel18}, even using the much smaller
proper motion sample of \citet{sohn16}.  In that paper, samples of
orbits constrained by the observations were obtained with and without
the use of proper motion.  Including proper motion pushed the derived
orbits into the tail of the distributions obtained when it was
omitted, indicating a significant level of tension.
Our much more extensive proper motion dataset renders the level of
disagreement far worse.

The inferred galactocentric velocity direction depends on the
motion of the sun.
If we allow some freedom in the solar motion,
as in orbit~2, we can improve the fit to the observables somewhat.
However, we find that good agreement between the track and an orbit
is only obtained for implausible choices of the solar motion,
requiring velocities differing by $>50 \kms$
from currently preferred values and
involving substantial motion out of the disk plane.
In orbit~3 we instead
force better agreement with the spatial track of the stream
(again see Figure~\ref{fig.misalign}),
at the cost of a $\tsim 30 \kms$ disagreement with the observed radial velocity
over a wide range in longitude.
In these various fits the pericenter of the orbit
is consistently about $17 \kpc$, as dictated by the
spatial track, 
but the apocenter and orbital period are not well constrained.  

It is not clear that a full dynamical model of the stream would 
necessarily reduce the misalignment we find here.
In dynamical simulations the leading part of a tidal stream tends to lie inside the orbit,
whereas the leading part of our observed track lies outside the fitted orbits.
Other effects such as dynamical friction, or interactions with massive satellites
such as Sagittarius or the LMC, may play a role in shaping the stream's path.
These effects are outside the scope of our preliminary investigation.

We note that we have tried various other potential forms,
including bulge-disk-halo models
with greater freedom in the halo properties, pure power-law potentials,
and flat or cored logarithmic halo halo potential, with
reasonable prolate or oblate halos.
Overall, the detailed parameterization turns out to make little difference to the fitted
orbital paths or the quality of the fits, perhaps because of the relatively
short portion of the orbit sampled by the stream.  
The one property of the potential that seems to be well constrained by the
fits is the circular velocity, or equivalently the radial acceleration or enclosed mass, 
in the range of radii probed by the stream (cf.\ \citealp{bonaca18}).  Not surprisingly,
we find circular velocity values similar to those suggested by Figure~\ref{fig.gcen}.

Although our main focus has been on the central track of the stream,
the widths of the stream in the coordinate dimensions also convey
useful physical information.  Constraining these widths is difficult
due to observational errors, contamination,
the likely non-Gaussian profiles of the stream,
and possible substructure
or other deviations from a smooth, constant-width stream.
We have thus not attempted a rigorous computation of any of the dispersions.
Nevertheless, over the
interval $0\degree < \Lambda < 50\degree$ at least, the stream appears
fairly narrow in all coordinates.  
In Section~\ref{sec.analysis} we estimated
$\sigma_{B} \approx 1.3 \degree$,
$\sigma_{v} \approx 7 \kms$, and
$\sigma_{DM} \loa  0.2$~mag.
We can compare these to the estimates of \citet{hendel18} based on a sample of RR Lyrae:
$\sigma_{B} \approx 1.0 \degree$,
$\sigma_{v} \approx 30 \kms$, and
$\sigma_{DM} \approx  0.22$~mag.
The samples are not entirely equivalent due to our
proper-motion cleaning, which eliminates some of the outliers,
and we have not chosen the same portion of the stream to estimate the widths.
Furthermore, the much larger velocity dispersion in the RR Lyrae sample
may be due in part to incomplete correction for the effects of radial pulsations.

\citet{hendel18} used a series of simulations matching their best-fit 
orbit and varying in progenitor mass to interpret their derived
dispersions.
(As these were one-component simulations,
this progenitor mass corresponds only to the
dense portion mixed with the stars that can survive the initial phases
of tidal stripping.)
The width in latitude pointed to a low progenitor mass
($<3 \times 10^7 \msun$), while the distance and velocity dispersions pointed
to a high mass, $\tsim 3 \times 10^9 \msun$.  To reconcile these, they
suggested that the latitude width apparent in maps of main-sequence
stars could be underestimated due to the background population.
A dispersion of $\approx 2.5 \degree$ would be required to be consistent
with their high preferred progenitor mass.

With proper motion selection,
the contamination we find in the central portion of the stream
(see Figure~\ref{fig.sky}) is low enough
that a dispersion of this size seems unlikely.
Instead, we suggest
that the small dispersions in latitude and velocity together with the trends
found by \citet{hendel18} point to a much lower progenitor
mass of $\tsim 10^8 \msun$.
Meanwhile, the distance dispersion is sensitive to the
assumed precision of observational distance estimates,
the exclusion of outliers,
and the longitude range used to compute the dispersion.
We would not be surprised by a true distance modulus dispersion
as low as $\tsim 0.1$~mag, consistent with a mass $\tsim 10^8 \msun$.

All these quantitative assessments of the progenitor are of course
quite imprecise, given the lack of an accurate and well-constrained 
model for the stream and the many possible sources of observational
or theoretical error.
Even so, the smaller mass we prefer may be easier to reconcile with
the low mean metallicities of the Orphan Stream stars.
For example, Draco, Ursa Minor, and Sextans were
the three dSph we used as proxies for the Orphan progenitor in
Section~\ref{sec.distance} due to their similar metallicity.
These have dynamical masses within the half-light radius of
only $10^7 \msun$, $10^7 \msun$, and $3 \times 10^7 \msun$
respectively \citep{alan12}.

\section{CONCLUSIONS}
\label{sec.conclusions}
We have used data from \Gaia\ DR2 and other surveys to constrain the properties of the Orphan Stream,
over the entire region where it has previously been detected.
We have clearly detected the proper motion of the stream over more than $90\degree$ in longitude.
This result verifies and greatly extends an earlier detection using \HST\ data \citep{sohn16}.
We confirm and slightly update the previously obtained tracks of distance and velocity,
with much lower ambiguity due to unrelated stars.
Although increased contamination near the disk makes detection more difficult,
we find evidence that the spatial path of the stream deviates from a great circle
by about $2\degree$ in that area.
Proper motion cleaning of previous spectroscopic surveys
narrows, but does not eliminate, the metallicity dispersion in the stream.
Consistent with this,
the stream component exhibits a narrow red giant branch in the color-magnitude diagram.
The low metallicity and small dispersions in the various
observational dimensions suggest a progenitor
of mass $\tsim 10^8 \msun$.

The motion of the stream suggests a circular velocity of about
$200$--$220 \kms$ at $30 \kpc$, consistent with results using other
tracers.  Otherwise, the inferred behavior of the stream is not
strongly sensitive to the form of the Galactic gravitational potential.
The stream path deviates in significant ways from the behavior of an
orbit; in particular, it does not point in the same direction as the
3d velocity vector.  
These apparent deviations can be reduced by shifts in the
assumed solar motion, but optimum agreement requires unreasonable
solar parameters.  More work will be needed to constrain the total
extent of the stream and understand its dynamics.  We anticipate
significant observational progress in the near future from further
data mining of \Gaia\ and other surveys, future \Gaia\ releases, and targeted
followup of stars along the stream.
\begin{table*}
\caption{
Orbits roughly matching the Orphan Stream.
Kinematic properties (latitude $B_0$, distance $d_0$, proper motion
in stream coordinates $\mu_{\Lambda,0}$ and $\mu_{B,0}$, 
and heliocentric radial velocity $v_0$)
are specified at longitude $\Lambda=20\degree$.
Orbits are integrated in a potential consisting of a 
Hernquist bulge with $M_b = 4.5 \times 10^9 \msun$ and $a_b = 0.44 \kpc$,
a Miyamoto-Nagai disk with $M_d = 6.8 \times 10^{10} \msun$,
$a_d=3.0 \kpc$, and $b_d=0.28 \kpc$,
and a Navarro-Frenk-White halo parameterized as
$\rho(r) = M_h / (4 \pi r_h^3) (r/r_h)^{-1} (1 + r/r_h)^{-2} $.
Solar motion parameters are relative to the Galactic center.
Here latitude is in degrees,
distance in $\kpc$, velocities in $\kms$, mass in $10^{11} \msun$,
and proper motion in $\masyr$.
}
\label{tab.orbits}
\begin{tabular}{rrrrrrrrrrr}
\hline
Orbit &
$B_0$&
$d_0$&
$\mu_{\Lambda,0}$&
$\mu_{B,0}$&
$v_0$&
$M_h$&
$r_h$&
$v_{\sun,x}$&
$v_{\sun,y}$&
$v_{\sun,z}$
\\
\hline
1&
$-0.181$&
$21.37$&
$-1.372$&
$-1.403$&
$242.1$&
$6.06$&
$16$&
$11.10$&
$245.24$&
$7.25$\\
2&
$-0.138$&
$20.63$&
$-1.371$&
$-1.384$&
$238.3$&
$4.98$&
$16$&
$11.57$&
$222.26$&
$13.96$\\
3&
$-0.131$&
$21.30$&
$-1.260$&
$-1.348$&
$247.0$&
$3.13$&
$16$&
$17.04$&
$223.78$&
$12.87$\\
\end{tabular}
\end{table*}
\section*{ACKNOWLEDGMENTS}
Support for this work was provided by NASA through grants for programs
GO-13443 and AR-15017
from the Space Telescope Science Institute (STScI), which is
operated by the Association of Universities for Research in Astronomy
(AURA), Inc., under NASA contract NAS5-26555.
This work has made use of data from the European Space Agency (ESA) mission
\Gaia\ (\url{https://www.cosmos.esa.int/gaia}), processed by the \Gaia\
Data Processing and Analysis Consortium (DPAC,
\url{https://www.cosmos.esa.int/web/gaia/dpac/consortium}). Funding for the DPAC
has been provided by national institutions, in particular the institutions
participating in the \Gaia\ Multilateral Agreement.
The Pan-STARRS1 Surveys (PS1) and the PS1 public science archive have
been made possible through contributions by the Institute for
Astronomy, the University of Hawaii, the Pan-STARRS Project Office,
the Max-Planck Society and its participating institutes, the Max
Planck Institute for Astronomy, Heidelberg and the Max Planck
Institute for Extraterrestrial Physics, Garching, The Johns Hopkins
University, Durham University, the University of Edinburgh, the
Queen's University Belfast, the Harvard-Smithsonian Center for
Astrophysics, the Las Cumbres Observatory Global Telescope Network
Incorporated, the National Central University of Taiwan, the Space
Telescope Science Institute, the National Aeronautics and Space
Administration under Grant No. NNX08AR22G issued through the Planetary
Science Division of the NASA Science Mission Directorate, the National
Science Foundation Grant No. AST-1238877, the University of Maryland,
Eotvos Lorand University (ELTE), the Los Alamos National Laboratory,
and the Gordon and Betty Moore Foundation.
This research made use of numerous open-source Python packages.  These include
{\tt numpy};
{\tt matplotlib};
{\tt astropy} \citep{astropy18}, for which we recognize the particular kinematics-specific
contributions of Adrian Price-Whelan and Erik Tollerud;
{\tt pygmmis}, by Peter Melchior;
{\tt sfdmap}, by Kyle Barbary;
and {\tt galpy}, by Jo Bovy.
We thank Vasily Belokurov and David Hendel for helpful conversations.
Preparation for this work was aided by the stimulating environment
at the 2018 NYC Gaia Sprint, hosted by the Center for Computational Astrophysics
of the Flatiron Institute in New York City.
This project is part of the HSTPROMO
(High-resolution Space Telescope PROper MOtion)
Collaboration\footnote{http://www.stsci.edu/$\sim$marel/hstpromo.html},
a set of projects aimed at improving our dynamical understanding of
stars, clusters and galaxies in the nearby Universe through
measurement and interpretation of proper motions from \HST, \Gaia, and
other space observatories. We thank the collaboration members for the
sharing of their ideas and software.
\footnotesize{
\bibliographystyle{mnras}
\bibliography{combined}
\label{lastpage}
}

\end{document}